\documentclass[aps,prl,twocolumn,superscriptaddress,showpacs,floatfix]{revtex4-1}
\usepackage{graphicx,amsmath}
\pdfoutput=1

\newcommand{\Startsubfig}[2]{Figure~\ref{fig:#1}(#2)}
\newcommand{\subfig}[2]{Fig.~\ref{fig:#1}(#2)}
\newcommand{\allfig}[1]{Fig.~\ref{fig:#1}}

\begin{document}

\title{Cotunneling drag effect in \protect{C}oulomb-coupled quantum dots}
\author{A. J. Keller}
	\altaffiliation{Present address: Institute for Quantum Information and Matter, California Institute of Technology, Pasadena, California 91125, USA}
	\affiliation{Department of Physics, Stanford University, Stanford, California 94305, USA}

\author{J. S. Lim}
	\affiliation{School of Physics, Korea Institute for Advanced Study, Seoul 130-722, Korea}

\author{David S\'anchez}
	\affiliation{IFISC (UIB-CSIC), E-07122 Palma de Mallorca, Spain}
	
\author{Rosa L\'opez}
	\affiliation{IFISC (UIB-CSIC), E-07122 Palma de Mallorca, Spain}

\author{S. Amasha}
	\altaffiliation{Present address: MIT Lincoln Laboratory, Lexington, Massachusetts 02420, USA}
	\affiliation{Department of Physics, Stanford University, Stanford, California 94305, USA}

\author{J. A. Katine}
	\affiliation{HGST, San Jose, CA 95135, USA}

\author{Hadas Shtrikman}
	\affiliation{Department of Condensed Matter Physics, Weizmann Institute of Science, Rehovot 96100, Israel}

\author{D. Goldhaber-Gordon}
	\email{goldhaber-gordon@stanford.edu}
	\affiliation{Department of Physics, Stanford University, Stanford, California 94305, USA}	

\begin{abstract}
In Coulomb drag, a current flowing in one conductor can induce a voltage across an adjacent conductor via the Coulomb interaction. The mechanisms yielding drag effects are not always understood, even though drag effects are sufficiently general to be seen in many low-dimensional systems. In this Letter, we observe Coulomb drag in a Coulomb-coupled double quantum dot (CC-DQD) and, through both experimental and theoretical arguments, identify cotunneling as essential to obtaining a correct qualitative understanding of the drag behavior. \end{abstract}

\maketitle

Coulomb-coupled quantum dots yield a model system for Coulomb drag~\cite{Sanchez2010}, the phenomenon where a current flowing in a so-called drive conductor induces a voltage across a nearby drag conductor via the Coulomb interaction~\cite{drag}. Though charge carriers being dragged along is an evocative image, as presented in early work on coupled 2D-3D~\cite{Solomon1989} or 2D-2D~\cite{Gramila1991} semiconductor systems, later measurements in graphene~\cite{Kim2011, Gorbachev2012}, quantum wires in semiconductor 2DEGs~\cite{Debray2001, Yamamoto2006, Laroche2011, Laroche2014}, and coupled double quantum dots~\cite{Shinkai2009} have indicated that the microscopic mechanisms leading to Coulomb drag can vary widely. For example, collective effects are important in 1D, but less so in other dimensions. All drag effects require interacting subsystems and vanish when both subsystems are in local equilibrium.

A perfect Coulomb drag with equal drive and drag currents has been observed in a bilayer 2D electron system: effectively a transformer operable at zero frequency~\cite{Nandi2012}. Coulomb-coupled quantum dots can rectify voltage fluctuations to unidirectional current, with possible energy harvesting applications~\cite{Hartmann2015,Thierschmann2015}. This rectification of nonequilibrium fluctuations is similar to a ratchet effect, as observed in charge-~\cite{Linke1999,Onac2006,Khrapai2006,Roche2015} and spin-based nanoelectronic devices~\cite{Costache2010}, as well as in rather different contexts such as suspended colloidal particles in asymmetric periodic potentials~\cite{Rousselet1994}. Coulomb-coupled dots have also been proposed as a means for testing fluctuation relations out of equilibrium~\cite{Sanchez2010}.

An open question is how higher-order tunneling events in the quantum coherent limit contribute to Coulomb drag processes~\cite{Bischoff2015}. In this Letter, we present experimental measurements and theoretical arguments showing that simultaneous tunneling of electrons (cotunneling) is crucial to describe drag effects qualitatively in Coulomb-coupled double quantum dots (CC-DQDs). Previous theoretical work has obtained drag effects with sequential tunneling models~\cite{Sanchez2010} (for an exception, see Ref.~\cite{kaa16}), and these models have been invoked in measurements of stacked graphene quantum dots~\cite{Bischoff2015}. We demonstrate here that for a DQD, cotunneling contributes to the drag current at the same order as sequential tunneling in a perturbation expansion. This has profound consequences in experiment, notably a measurable drag current even when the drag dot is far off resonance, and a gate voltage-dependent vanishing of the Coulomb gap above which drag current can be measured. Our experiment shows that the drag mechanisms considered can be observed in highly tunable GaAs/AlGaAs QDs, not only in graphene. We also achieve the unexplored regime $kT \ll \hbar \Gamma$, where $T$ is temperature and $\Gamma$ is a tunnel rate, which is outside the scope of theories to date.

\begin{figure}
\begin{center}
\includegraphics{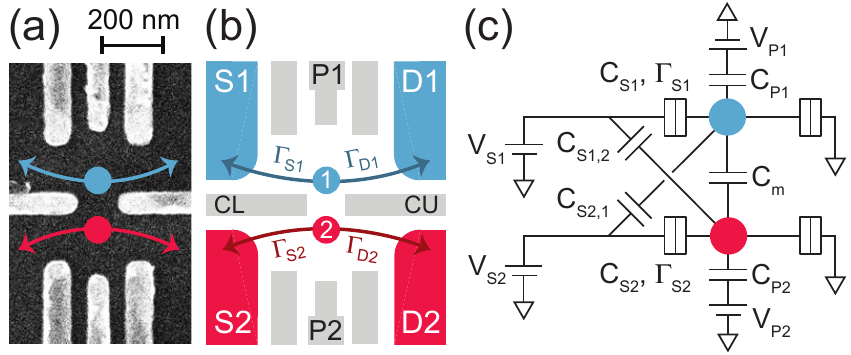}
\end{center}
\caption{Device and model. (a) Top-down SEM image of a device nominally identical to that measured. Ti/Au gate electrodes (light gray) are patterned on the substrate surface (dark gray). Colored circles represent the QDs. Arrows indicate where electrons can tunnel. (b) Cartoon showing names of gates, reservoirs, and dots. $\Gamma_{Si}$ is the tunnel rate between reservoir $Si$ and dot $i$. (c) Capacitor and tunnel junction network. Interdot tunneling is strongly suppressed and not included in the model. Direct capacitance between gate P1 (2) and dot 2 (1) is omitted from the diagram for clarity, along with some labels.
\label{fig:device}}
\end{figure}

Our device (\subfig{device}{a}) consists of a lithographically-patterned AlGaAs/GaAs heterostructure with electron density $2 \times 10^{11}~\mbox{cm}^{-2}$ and mobility $2 \times 10^6~\mbox{cm}^{2}/\mbox{Vs}$. All measurements are taken in a dilution refrigerator. The interdot tunnel rate is made negligible, tens of times smaller than all other dot-lead tunnel rates, by applying appropriate voltages on gate electrodes named CL and CU (\subfig{device}{b}), as done previously with the very same device ~\cite{Amasha2013:Pseudospin,Keller2014:EmergentSU4}. The device then realizes a capacitance and tunnel junction network sufficient to observe Coulomb drag (\subfig{device}{c})~\cite{Sanchez2010}. We measure $G_i=dI_i/dV_{Si}$ and $I_i$ for dot $i \in \{1,2\}$, using standard current preamp+lock-in amplifier techniques. The near-DC current measurements of $I_i$ were obtained by filtering current amplifier outputs with single-stage low-pass filters (R=2.7~k$\Omega$, C=10~$\mu$F). In all measurements we present in this paper, an in-plane field of 2.0 T and an out-of-plane field of 0.1 T were applied. The application of a small out-of-plane field can help tune couplings. The large in-plane field breaks spin degeneracy of the dot levels to simplify the discussion. The magnetic field is not necessary to observe drag currents.

\begin{figure}
\begin{center}
\includegraphics[width=3.4in]{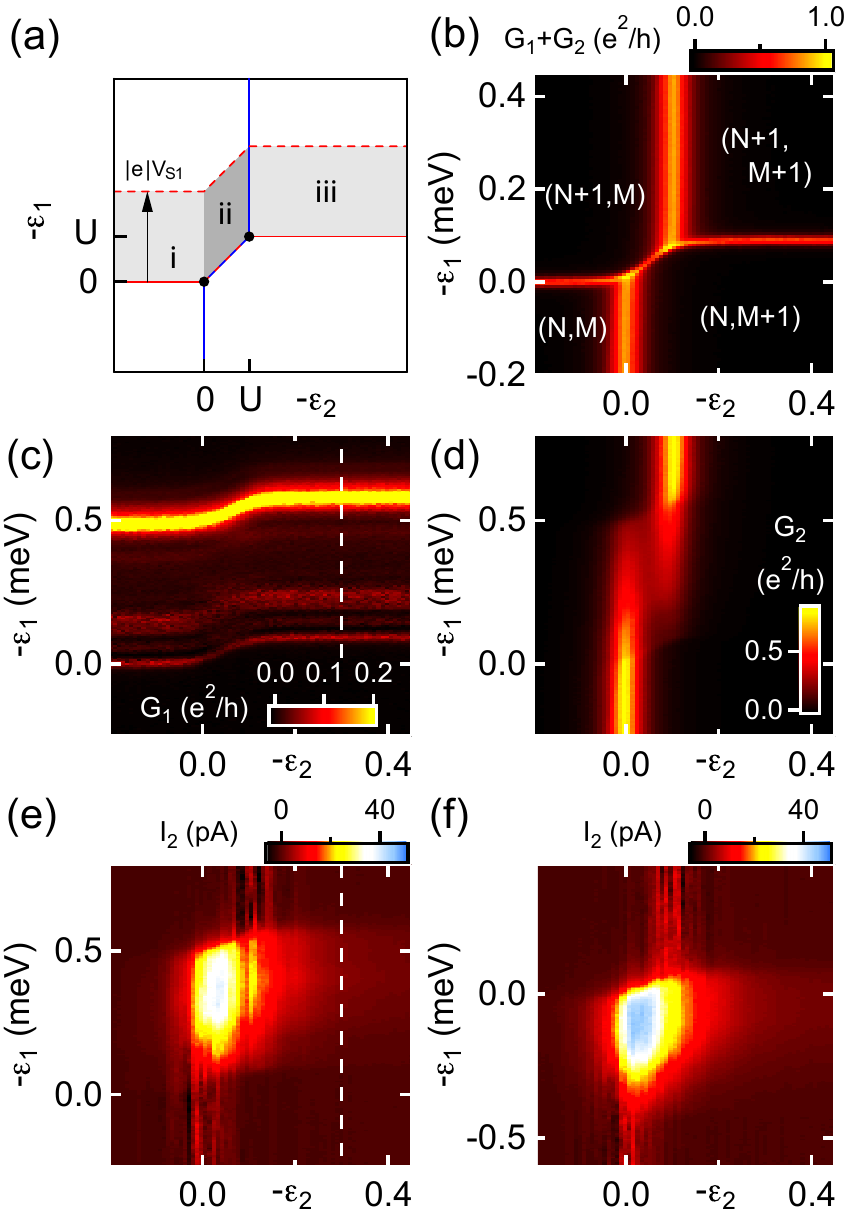}
\end{center}
\caption{Coulomb drag. (a) Schematic charge stability diagram for a CC-DQD. Dots indicate triple points. Red (blue) solid lines are charge transitions for dot 1 (2). As $V_{S1}$ increases from zero, excited states appear in $G_1$ within shaded regions. Roman numerals are used later for reference. (b) Sum of measured conductances $G_i = dI_i / dV_{Si}$ for $V_{S1} = V_{S2} = 0$, as a function of dot levels $\varepsilon_1, \varepsilon_2$.
(c,d) Measured $G_1$ (c) and $G_2$ (d) for $V_{S1} = 0.5$~mV. (e,f) Measured $I_2$ for $V_{S1} = 0.5$~mV (e) and $V_{S1} = -0.5$~mV (f). In both cases the current $I_2$ flows in the same direction, is strongest in region (ii), and persists in regions (i) and (iii). Dashed white lines in (c) and (e) are discussed in the text.
\label{fig:drag}}
\end{figure}

For zero source-drain bias, peaks in measured $G_i = dI_i/dV_{Si}$ correspond to charge transitions of the dots (\subfig{drag}{a}).  The measured, summed conductance $G_1+G_2$ shows both charge transitions (\subfig{drag}{b}). By a change of basis from the gate voltage axes $V_{P1}$ and $V_{P2}$, we measure along the dot level axes $-\varepsilon_1$ and $-\varepsilon_2$. The dots can be Coulomb blockaded as both temperature $T$ and the dot-lead tunnel rates $\Gamma_i = \Gamma_{S,i}+\Gamma_{D,i}$ are small compared to the addition energies $U_i$. The numbers of electrons on the dots are unknown in this experiment, but we can label how many there are relative to some $(N,M)$ in \subfig{drag}{b}. By taking horizontal or vertical cuts on the bottom or left edges of \subfig{drag}{b} respectively, we extract the FWHM of the observed peaks and find $\Gamma_1 = 15$~$\mu$eV and $\Gamma_2 = 47$~$\mu$eV, considerably larger than $T = 20$~mK~$\approx 1.7$~$\mu$eV. Quantum coherent processes may therefore be important.

When applying a source-drain bias $V_{S1}$ ($V_{D1}$ is fixed at zero), a window in $-\varepsilon_1$ should open wherein peaks in $G_1$, reflecting excited states of dot 1, may be observed (\subfig{drag}{a}). The location of this window depends on $-\varepsilon_2$; we define three regions to aid in discussion. In \subfig{drag}{c}, we apply $V_{S1} = 0.5$~mV and see excited states, e.g. between $-\varepsilon_1 = 0$ and $|e|V_{S1}$ in region (i), or between $-\varepsilon_1=U$ and $|e|V_{S1}+U$ in region (iii), where $U$ is the interdot charging energy~\cite{biasnote}. For $\varepsilon_{1},\varepsilon_{2}$ within any shaded region of \subfig{drag}{a}, the measured $G_1$ is accompanied by a non-zero DC current $I_1$ that can drive Coulomb drag.

Keeping $V_{S1} = 0.5$~mV, and noting that both reservoirs S2 and D2 are grounded, we easily resolve a drag current $I_2 \sim 40$~pA in region (ii) (\subfig{drag}{e}). More surprisingly, we still see significant $I_2$ in region (iii), where sequential tunneling in dot 2 should be very suppressed, with the current decreasing as $-\varepsilon_2$ grows. Current on the order of 0.5 pA is also measured in region (i), decreasing as $-\varepsilon_2$ decreases. $G_2$ is apparently insensitive to the current flowing in dot 1 in regions (i,iii) (\subfig{drag}{d}). Upon inverting the sign of $V_{S1}$, we observe qualitatively similar features in $I_2$ (\subfig{drag}{f}). The drag current flows in the same direction, regardless of $V_{S1}$'s sign. Vertical cuts in \subfig{drag}{c,e} are compared in Sec. E of Ref.~\cite{SI} and indicate sensitivity of $I_2$ to dot 1's excited states.

\begin{figure}
\begin{center}
\includegraphics{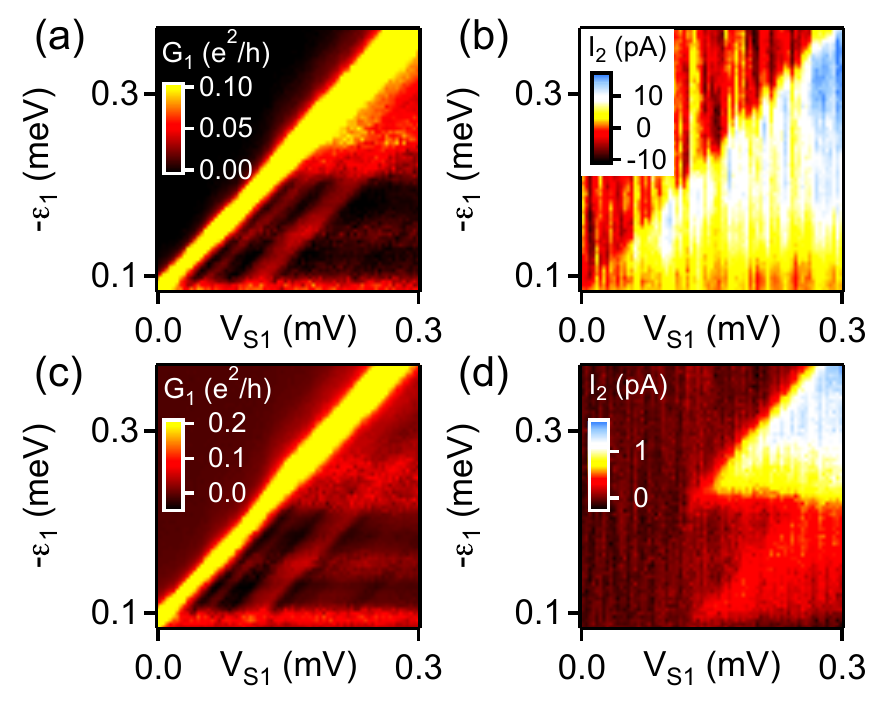}
\end{center}
\caption{For small drive bias $V_{S1}$, the drag current $I_2$ appears to vanish depending on the drag dot's level. (a) $G_1$ for $-\varepsilon_2 = 0.12$~meV, on the border of region (ii) and (iii) in \subfig{drag}{a}. The color scale is saturated to emphasize fine features.
(b) Drag current $I_2$ for $-\varepsilon_2 = 0.12$~meV persists even in the limit that drive bias $V_{S1} \rightarrow 0$.
(c) $G_1$ for $-\varepsilon_2 = 0.47$~meV, well within region (iii) in \subfig{drag}{a}. The color scale is saturated, and appears similar to (a).
(d) $I_2$ for $-\varepsilon_2 = 0.47$~meV (region (iii)). Below $V_{S1} \sim 0.12$~meV, the drag current is unmeasurable. This gap also appears for $-\varepsilon_1 < 0.1$ and negative $V_{S1}$ (not shown), and is the same value within measurement accuracy.
\label{fig:gap}}
\end{figure}

Having demonstrated Coulomb drag, we perform bias spectroscopy (\allfig{gap}) to detect the presence of a Coulomb gap---an energy below which drag currents are vanishing---as indicated in prior theoretical studies of drag in CC-DQDs~\cite{Sanchez2010}. For $-\varepsilon_2$ on the border of region (ii) and (iii) of \subfig{drag}{a}, such a gap does not clearly appear. \Startsubfig{gap}{a} and \ref{fig:gap}(b) show $G_1$ and $I_2$ respectively, and a non-zero current $I_2$ flows provided $0.1 < -\varepsilon_1 < 0.1+|e|V_{S1}$. Current noise is intrinsically strong for this tuning of $-\varepsilon_2$, as also seen in \subfig{drag}{e,f}. However, if $-\varepsilon_2$ is well within region (iii) of \subfig{drag}{a}, there appears to be a gap. (\subfig{gap}{c,d}). Though $G_1$ looks similar to before, $I_2$ looks dramatically different, with much less current noise, smaller average drag currents, and a gap of $\sim 0.12$~meV. The range of ($V_{S1}$, $-\varepsilon_1$) where drag current flows appears to be bounded by excited states seen in \subfig{gap}{c}. The size of the gap does not seem to depend on $-\varepsilon_2$ in region (iii); we have verified this for $-\varepsilon_2 \in \{0.21, 0.29,0.38\}$. At each of these values, $I_2$ looks much like it does in \subfig{gap}{d}, but with different magnitude. We note the observed gap of 0.12~meV is close to $U \sim 0.1$~meV.

\begin{figure}
\begin{center}
\includegraphics{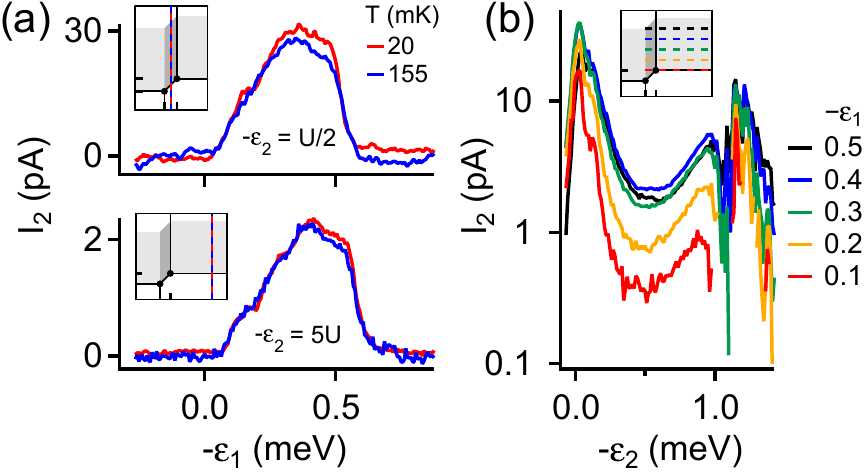}
\end{center}
\caption{
Temperature and dot level dependence of drag current. Icons indicate where cuts are taken (as in \subfig{drag}{a}). $V_{S1} = 500$~$\mu$V, $V_{S2} = 0$. (a) Temperature-dependent $I_2$. Top: $-\varepsilon_2 = 0.05$~meV = $U/2$ (middle of region (ii) in \subfig{drag}{a}); bottom: $-\varepsilon_2 = 0.5$~meV = $5U$ (deep in region (iii)). The drag current does not change appreciably from 20 to 155~mK in either case.
(b) Drag current $I_2$ can be measured even when dot 2's levels are far off resonance, provided a current is flowing in dot 1.
\label{fig:deps}}
\end{figure}

To elucidate the mechanisms behind the Coulomb drag, we study the $T$ and $-\varepsilon_2$ dependence of $I_2$. Changing the temperature has a weak effect if any in the range 20 to 155~mK, in both regions (ii) (\subfig{deps}{a}; top) and (iii) (\subfig{deps}{a}; bottom). Our electron temperature determination is based on calibrating a ruthenium-oxide resistive thermometer in the mixing chamber of our dilution refrigerator to Coulomb blockade thermometry measurements performed in equilibrium. As such, we cannot rule out the possibility that our base electron temperature is higher than 20~mK in the presence of large biases. The $-\varepsilon_2$ dependence shows that a drag current is measurably large for {\em any} value of $-\varepsilon_2$ (\subfig{deps}{b}). For small biases, prior theories with sequential tunneling only~\cite{Sanchez2010} would yield vanishingly small drag currents if the dot levels were off resonance by more than the width of the Fermi-Dirac distribution.

\begin{figure}
\begin{center}
\includegraphics[width=0.46\textwidth]{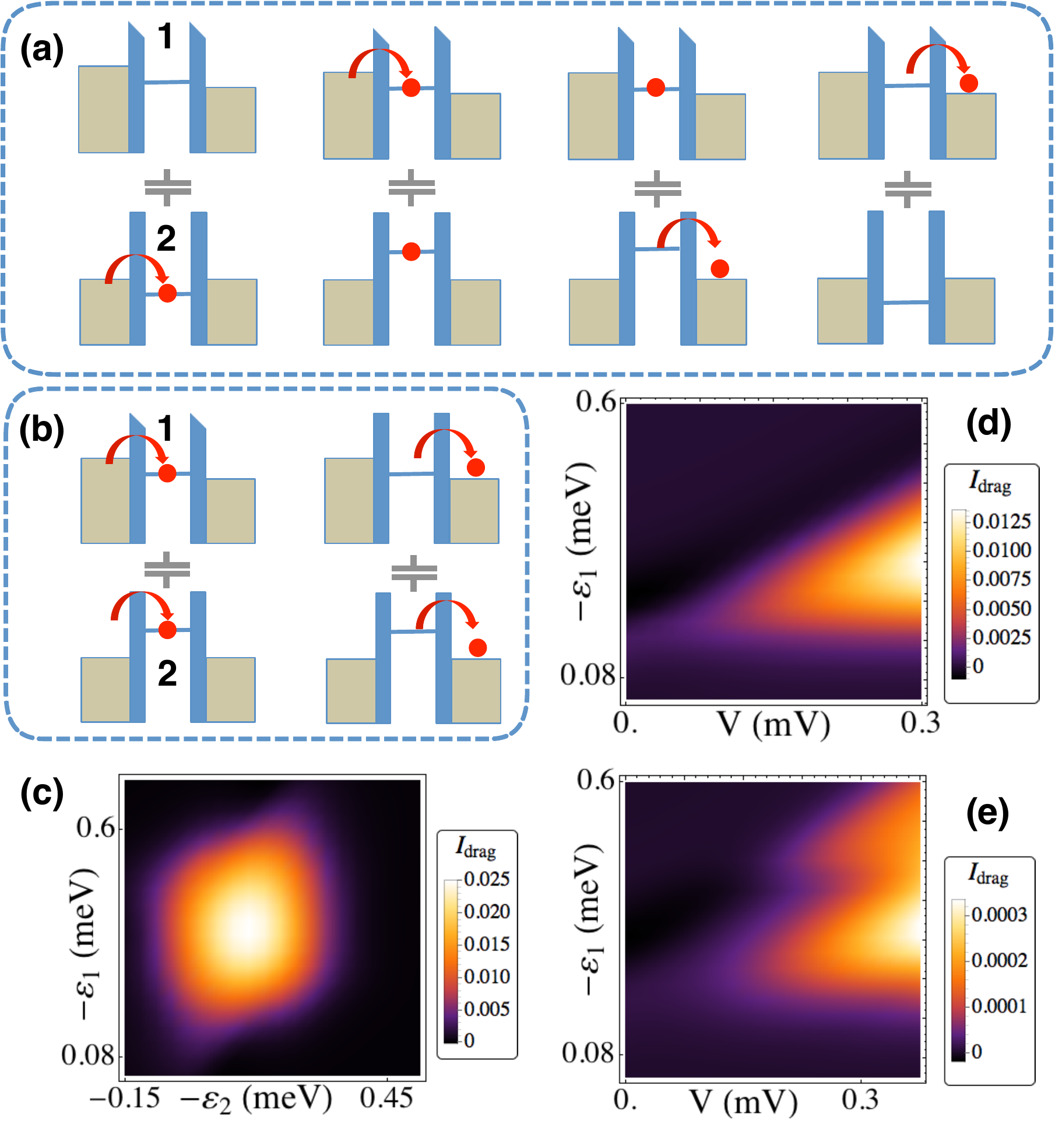}
\end{center}
\caption{Theory of cotunneling leading to drag.
(a) Cartoon of a sequential process leading to drag current. An electron hops into the drag dot (dot 2). Next, an electron hops into the drive dot (dot 1), causing the dot 2 level to rise due to interdot Coulomb repulsion in turn allowing the electron in 2 to be transferred to the right. The whole process involves four tunneling rates and is hence of order $\Gamma^4$. (b) Pure cotunneling process leading to drag current. Two electrons tunnel simultaneously onto the dots. They then tunnel off simultaneously. Since each cotunneling process has a probability $\Gamma^2$, the cotunneling drag process is of order $\Gamma^4$ as in (a). (c) Calculated drag current in units of $e\Gamma/\hbar$ for drive voltage $V=0.5$~mV and system parameters extracted from the experiment. (d,e) Drag current as a function of drive dot level and $V$ for (d) $\varepsilon_2=-0.2$~meV and (e) $\varepsilon_2=-0.4$~meV.
\label{fig:theor}}
\end{figure}

We now show that the interpretation of our data is compatible with a theoretical model that includes both sequential and cotunneling processes. Remarkably, we find here that sequential and cotunneling processes contribute to the drag current to the same order despite the cotunneling rate being calculated from a higher-order perturbative term. This is illustrated in Fig.~\ref{fig:theor}: while a sequential drag current needs four hoppings in four steps (Fig.~\ref{fig:theor}(a)), a pure cotunneling current requires only two steps (Fig.~\ref{fig:theor}(b)). Therefore, a complete theory of the drag effect in CC-DQD must take into account both types of processes on equal footing. We discuss our results on the basis of a master equation approach. From Fig.~\ref{fig:drag}(b) we consider four charge states in the CC-DQD system: $\{|0\rangle=|00\rangle\,,|1\rangle=|10\rangle\,,|2\rangle=|01\rangle\,,|d\rangle=|11\rangle\}$. The set of stationary probabilities that the system is in any of these states obeys the kinetic equations
$0=\mathbf{\Gamma} \boldsymbol{p}$, where $\boldsymbol{p}=(p_0,p_1,p_2,p_d)^T$ fulfills $p_0+p_1+p_2+p_d=1$ and $\mathbf{\Gamma}$ denotes the matrix containing the rates. A representative equation reads
\begin{equation}
0=\Gamma_{10}p_1+\Gamma_{20} p_2+\gamma_{d0}p_0-(\Gamma_{02}+\Gamma_{01}+\gamma_{0d})p_0\,.
\end{equation}
(The remaining equations are shown in Sec.~D of Ref.~\cite{SI}.) Here, $\Gamma_{0i(i0)}=\sum_\alpha \Gamma_{0i(i0)}^{\alpha i}$
and $\gamma_{0d(d0)}=\sum_{\alpha,\beta} \gamma_{0d(d0)}^{\alpha 1\beta 2}$.
$\Gamma_{0i(i0)}^{\alpha i}$ is the sequential rate that describes the addition (removal) of an electron into (from) dot $i=1,2$ from (to) lead $\alpha i$ with $\alpha=S,D$
and $\gamma_{0d(d0)}^{\alpha 1\beta 2}$ is the cotunneling rate that characterizes the simultaneous
tunneling of two electrons on (off) the CC-DQD with $\beta=S,D$.
The expressions for these rates follow from a perturbation expansion in the tunneling coupling~\cite{bruus},
valid for $k T> \Gamma$. To lowest order one finds $\Gamma_{0i}^{\alpha i}=(\Gamma_{\alpha i}/\hbar) f_{\alpha i}(\mu_i)$ and $\Gamma_{i0}^{\alpha i}=(\Gamma_{\alpha i}/\hbar) [1-f_{\alpha i}(\mu_i)]$
with $\Gamma_{\alpha i}$ the level broadening of dot $i$ due to hybridization with lead $\alpha i$,
$f_{\alpha i}(x)=1/[1+e^{(x-\mu_{\alpha_i})/k T}]$ the Fermi-Dirac distribution function
($\mu_{\alpha_i}=E_F+eV_{\alpha i}$) and $\mu_i$ the electrochemical potential of dot $i$.
This has to be determined from an electrostatic model that takes into account both the polarization charges due to electric shifts in the leads [$C_{S,i}$ in Fig.~\ref{fig:device}(c)] and the interdot electron-electron interaction
[$C_m$ in  Fig.~\ref{fig:device}(c)]. The cotunneling rates (Fig.~\ref{fig:theor}(b)) are found in the next order in the tunneling coupling,
\begin{align}\label{eq_gamma0d}
\gamma_{0d}^{\alpha\bar{i} \beta i}&=\frac{2\pi}{\hbar}
\int d\varepsilon \left|
\frac{t_{\alpha i}^0 t_{\beta i}^1}{\varepsilon-\mu_{\bar{i}}+i\eta}
-\frac{t_{\alpha\bar{i}}^1 t_{\beta i}^0}{\varepsilon-\mu_{\bar{i}}-U+i\eta}
\right|^2\nonumber\\
&\times \rho_{\alpha{\bar{i}}}\rho_{\beta i} f_{\alpha{\bar{i}}}(\varepsilon) f_{\beta i}(\mu_i+\mu_{\bar{i}}+U-\varepsilon)\,,
\end{align}
where $\bar{i}=2(1)$ if $i=1(2)$, $U$ is given by a combination of the system capacitances, $\rho$ is the lead density of states, and a small imaginary part $\eta\to 0^+$ is added to avoid the divergence due to the infinite lifetime of the virtual intermediate states~\cite{averin,turek}. $\gamma_{d0}^{\alpha\bar{i} \beta i}$ is found by replacing $f$ with $(1-f)$ in Eq.~\eqref{eq_gamma0d}. Importantly, $t_{\alpha \bar{i}}^0$ ($t_{\beta i}^1$) is the tunneling amplitude for barrier $\alpha \bar{i}$ (${\beta i}$) when zero (one) charges are present in the DQD. That the amplitudes depend on the charge state derives physically from the general fact that tunneling is energy dependent and the dot levels shift with the charge state according to the electrostatic model. This is a crucial condition to generate drag currents. The probability that the sequence $|0\rangle\to|2\rangle\to|d\rangle\to|1\rangle\to|0\rangle$ drags an electron from left to right [Fig.~\ref{fig:theor}(a)] must differ from the reverse sequence. This occurs only if $\Gamma$ is energy dependent, for both sequential~\cite{Sanchez2010} and cotunneling processes.

The drag current is given by
$I_{\rm drag}\equiv I_{S2}=e[\Gamma_{20}^{S2} p_2+\Gamma_{d1}^{S2} p_d-\Gamma_{02}^{S2} p_0-\Gamma_{d1}^{S2} p_1
+\sum_\alpha \gamma_{21}^{\alpha 1 S2} p_2
+\sum_\alpha \gamma_{d0}^{\alpha 1 S2} p_d
-\sum_\alpha \gamma_{0d}^{S2 \alpha 1} p_0
-\sum_\alpha \gamma_{12}^{S2\alpha 1} p_1
]$ (we take $\mu_{S2}=\mu_{D2}$). We extract the parameters from the experiment and plot the results in Fig.~\ref{fig:theor}(c) for a drive voltage $V=0.5$~mV~\cite{note}. Comparing with the data in Fig.~\ref{fig:drag}(e) we obtain a good agreement. We find in Fig.~\ref{fig:theor}(c) an extended region of nonzero drag current as compared to the sequential case~\cite{Sanchez2010}, although the size of this region observed in Fig.~\ref{fig:drag}(e) is even larger than predicted, probably due to increased coherence in the experiment at lower $T$.

In Fig.~\ref{fig:gap}(b) we saw no Coulomb gap. The theoretical dependence of $I_{\rm drag}$ with $V$ (Fig.~\ref{fig:theor}(d)) reproduces this observation, in stark contrast to the theory of Ref.~\cite{Sanchez2010}, further emphasizing the role of cotunneling.
Physically, transport can occur via nonlocal cotunneling processes ($|1\rangle\to|2\rangle$ or vice versa) without traversing the doubly occupied state $|d\rangle$, so the Coulomb gap disappears.
For a larger value of $|\varepsilon_2|$ the gap reappears (Fig.~\ref{fig:theor}(e)) in agreement with the experiment (Fig.~\ref{fig:gap}(d)).

In conclusion, cotunneling is essential to understanding drag effects in CC-DQDs. We extend the existing theoretical framework to account for cotunneling processes, which cannot be justifiably neglected, as seen in experimental data. Though the theoretical framework is only valid for high temperatures, we are encouraged by the qualitative agreement between experiment and theory. Explaining some features in the experiment---namely the apparently weak temperature dependence and the role of excited states---will require additional theory. Double quantum dots are a popular model system for many-body physics, and play important roles in quantum information.  Understanding the subtle transport mechanisms in double quantum dots may thus have broad implications.

\begin{acknowledgments}
	We are grateful to L. Peeters for discussions. This work was supported by the Gordon and Betty Moore Foundation grant no. GBMF3429, the U.S.-Israel BSF grant Nos. 2014014 \& 2008149, the NSF under DMR-0906062,  and the MINECO grant No.\ FIS2014-52564. A.J.K. acknowledges an ABB Stanford Graduate Fellowship and an IQIM Postdoctoral Scholarship from the Institute for Quantum Information and Matter, an NSF Physics Frontiers Center (NSF Grant No. PHY-1125565).
\end{acknowledgments}

\bibliography{drag}

\newcommand{\noopsort}[1]{} \newcommand{\printfirst}[2]{#1}
  \newcommand{\singleletter}[1]{#1} \newcommand{\switchargs}[2]{#2#1}
\begin{thebibliography}{31}%
\makeatletter
\providecommand \@ifxundefined [1]{%
 \@ifx{#1\undefined}
}%
\providecommand \@ifnum [1]{%
 \ifnum #1\expandafter \@firstoftwo
 \else \expandafter \@secondoftwo
 \fi
}%
\providecommand \@ifx [1]{%
 \ifx #1\expandafter \@firstoftwo
 \else \expandafter \@secondoftwo
 \fi
}%
\providecommand \natexlab [1]{#1}%
\providecommand \enquote  [1]{``#1''}%
\providecommand \bibnamefont  [1]{#1}%
\providecommand \bibfnamefont [1]{#1}%
\providecommand \citenamefont [1]{#1}%
\providecommand \href@noop [0]{\@secondoftwo}%
\providecommand \href [0]{\begingroup \@sanitize@url \@href}%
\providecommand \@href[1]{\@@startlink{#1}\@@href}%
\providecommand \@@href[1]{\endgroup#1\@@endlink}%
\providecommand \@sanitize@url [0]{\catcode `\\12\catcode `\$12\catcode
  `\&12\catcode `\#12\catcode `\^12\catcode `\_12\catcode `\%12\relax}%
\providecommand \@@startlink[1]{}%
\providecommand \@@endlink[0]{}%
\providecommand \url  [0]{\begingroup\@sanitize@url \@url }%
\providecommand \@url [1]{\endgroup\@href {#1}{\urlprefix }}%
\providecommand \urlprefix  [0]{URL }%
\providecommand \Eprint [0]{\href }%
\providecommand \doibase [0]{http://dx.doi.org/}%
\providecommand \selectlanguage [0]{\@gobble}%
\providecommand \bibinfo  [0]{\@secondoftwo}%
\providecommand \bibfield  [0]{\@secondoftwo}%
\providecommand \translation [1]{[#1]}%
\providecommand \BibitemOpen [0]{}%
\providecommand \bibitemStop [0]{}%
\providecommand \bibitemNoStop [0]{.\EOS\space}%
\providecommand \EOS [0]{\spacefactor3000\relax}%
\providecommand \BibitemShut  [1]{\csname bibitem#1\endcsname}%
\let\auto@bib@innerbib\@empty
\bibitem [{\citenamefont {S\'anchez}\ \emph {et~al.}(2010)\citenamefont
  {S\'anchez}, \citenamefont {L\'opez}, \citenamefont {S\'anchez},\ and\
  \citenamefont {B\"uttiker}}]{Sanchez2010}%
  \BibitemOpen
  \bibfield  {author} {\bibinfo {author} {\bibfnamefont {R.}~\bibnamefont
  {S\'anchez}}, \bibinfo {author} {\bibfnamefont {R.}~\bibnamefont {L\'opez}},
  \bibinfo {author} {\bibfnamefont {D.}~\bibnamefont {S\'anchez}}, \ and\
  \bibinfo {author} {\bibfnamefont {M.}~\bibnamefont {B\"uttiker}},\ }\href
  {\doibase 10.1103/PhysRevLett.104.076801} {\bibfield  {journal} {\bibinfo
  {journal} {Phys. Rev. Lett.}\ }\textbf {\bibinfo {volume} {104}},\ \bibinfo
  {pages} {076801} (\bibinfo {year} {2010})}\BibitemShut {NoStop}%
\bibitem [{\citenamefont {Narozhny}\ and\ \citenamefont
  {Levchenko}(2016)}]{drag}%
  \BibitemOpen
  \bibfield  {author} {\bibinfo {author} {\bibfnamefont {B.~N.}\ \bibnamefont
  {Narozhny}}\ and\ \bibinfo {author} {\bibfnamefont {A.}~\bibnamefont
  {Levchenko}},\ }\href@noop {} {\bibfield  {journal} {\bibinfo  {journal}
  {Rev. Mod. Phys.}\ }\textbf {\bibinfo {volume} {88}},\ \bibinfo {pages}
  {025003} (\bibinfo {year} {2016})}\BibitemShut {NoStop}%
\bibitem [{\citenamefont {Solomon}\ \emph {et~al.}(1989)\citenamefont
  {Solomon}, \citenamefont {Price}, \citenamefont {Frank},\ and\ \citenamefont
  {{La Tulipe}}}]{Solomon1989}%
  \BibitemOpen
  \bibfield  {author} {\bibinfo {author} {\bibfnamefont {P.~M.}\ \bibnamefont
  {Solomon}}, \bibinfo {author} {\bibfnamefont {P.~J.}\ \bibnamefont {Price}},
  \bibinfo {author} {\bibfnamefont {D.~J.}\ \bibnamefont {Frank}}, \ and\
  \bibinfo {author} {\bibfnamefont {D.~C.}\ \bibnamefont {{La Tulipe}}},\
  }\href@noop {} {\bibfield  {journal} {\bibinfo  {journal} {Phys. Rev. Lett.}\
  }\textbf {\bibinfo {volume} {63}},\ \bibinfo {pages} {2508} (\bibinfo {year}
  {1989})}\BibitemShut {NoStop}%
\bibitem [{\citenamefont {Gramila}\ \emph {et~al.}(1991)\citenamefont
  {Gramila}, \citenamefont {Eisenstein}, \citenamefont {MacDonald},
  \citenamefont {Pfeiffer},\ and\ \citenamefont {West}}]{Gramila1991}%
  \BibitemOpen
  \bibfield  {author} {\bibinfo {author} {\bibfnamefont {T.~J.}\ \bibnamefont
  {Gramila}}, \bibinfo {author} {\bibfnamefont {J.~P.}\ \bibnamefont
  {Eisenstein}}, \bibinfo {author} {\bibfnamefont {A.~H.}\ \bibnamefont
  {MacDonald}}, \bibinfo {author} {\bibfnamefont {L.~N.}\ \bibnamefont
  {Pfeiffer}}, \ and\ \bibinfo {author} {\bibfnamefont {K.~W.}\ \bibnamefont
  {West}},\ }\href {\doibase 10.1103/PhysRevLett.66.1216} {\bibfield  {journal}
  {\bibinfo  {journal} {Phys. Rev. Lett.}\ }\textbf {\bibinfo {volume} {66}},\
  \bibinfo {pages} {1216} (\bibinfo {year} {1991})}\BibitemShut {NoStop}%
\bibitem [{\citenamefont {Kim}\ \emph {et~al.}(2011)\citenamefont {Kim},
  \citenamefont {Jo}, \citenamefont {Nah}, \citenamefont {Yao}, \citenamefont
  {Banerjee},\ and\ \citenamefont {Tutuc}}]{Kim2011}%
  \BibitemOpen
  \bibfield  {author} {\bibinfo {author} {\bibfnamefont {S.}~\bibnamefont
  {Kim}}, \bibinfo {author} {\bibfnamefont {I.}~\bibnamefont {Jo}}, \bibinfo
  {author} {\bibfnamefont {J.}~\bibnamefont {Nah}}, \bibinfo {author}
  {\bibfnamefont {Z.}~\bibnamefont {Yao}}, \bibinfo {author} {\bibfnamefont
  {S.~K.}\ \bibnamefont {Banerjee}}, \ and\ \bibinfo {author} {\bibfnamefont
  {E.}~\bibnamefont {Tutuc}},\ }\href {\doibase 10.1103/PhysRevB.83.161401}
  {\bibfield  {journal} {\bibinfo  {journal} {Phys. Rev. B}\ }\textbf {\bibinfo
  {volume} {83}},\ \bibinfo {pages} {161401} (\bibinfo {year}
  {2011})}\BibitemShut {NoStop}%
\bibitem [{\citenamefont {Gorbachev}\ \emph {et~al.}(2012)\citenamefont
  {Gorbachev}, \citenamefont {Geim}, \citenamefont {Katsnelson}, \citenamefont
  {Novoselov}, \citenamefont {Tudorovskiy}, \citenamefont {Grigorieva},
  \citenamefont {MacDonald}, \citenamefont {Morozov}, \citenamefont {Watanabe},
  \citenamefont {Taniguchi},\ and\ \citenamefont
  {Ponomarenko}}]{Gorbachev2012}%
  \BibitemOpen
  \bibfield  {author} {\bibinfo {author} {\bibfnamefont {R.~V.}\ \bibnamefont
  {Gorbachev}}, \bibinfo {author} {\bibfnamefont {A.~K.}\ \bibnamefont {Geim}},
  \bibinfo {author} {\bibfnamefont {M.~I.}\ \bibnamefont {Katsnelson}},
  \bibinfo {author} {\bibfnamefont {K.~S.}\ \bibnamefont {Novoselov}}, \bibinfo
  {author} {\bibfnamefont {T.}~\bibnamefont {Tudorovskiy}}, \bibinfo {author}
  {\bibfnamefont {I.~V.}\ \bibnamefont {Grigorieva}}, \bibinfo {author}
  {\bibfnamefont {A.~H.}\ \bibnamefont {MacDonald}}, \bibinfo {author}
  {\bibfnamefont {S.~V.}\ \bibnamefont {Morozov}}, \bibinfo {author}
  {\bibfnamefont {K.}~\bibnamefont {Watanabe}}, \bibinfo {author}
  {\bibfnamefont {T.}~\bibnamefont {Taniguchi}}, \ and\ \bibinfo {author}
  {\bibfnamefont {L.~A.}\ \bibnamefont {Ponomarenko}},\ }\href {\doibase
  10.1038/nphys2441} {\bibfield  {journal} {\bibinfo  {journal} {Nature Phys.}\
  }\textbf {\bibinfo {volume} {8}},\ \bibinfo {pages} {896} (\bibinfo {year}
  {2012})}\BibitemShut {NoStop}%
\bibitem [{\citenamefont {Debray}\ \emph {et~al.}(2001)\citenamefont {Debray},
  \citenamefont {Zverev}, \citenamefont {Raichev}, \citenamefont {Klesse},
  \citenamefont {Vasilopoulos},\ and\ \citenamefont {Newrock}}]{Debray2001}%
  \BibitemOpen
  \bibfield  {author} {\bibinfo {author} {\bibfnamefont {P.}~\bibnamefont
  {Debray}}, \bibinfo {author} {\bibfnamefont {V.}~\bibnamefont {Zverev}},
  \bibinfo {author} {\bibfnamefont {O.}~\bibnamefont {Raichev}}, \bibinfo
  {author} {\bibfnamefont {R.}~\bibnamefont {Klesse}}, \bibinfo {author}
  {\bibfnamefont {P.}~\bibnamefont {Vasilopoulos}}, \ and\ \bibinfo {author}
  {\bibfnamefont {R.~S.}\ \bibnamefont {Newrock}},\ }\href@noop {} {\bibfield
  {journal} {\bibinfo  {journal} {J. Phys.: Condens. Matter}\ }\textbf
  {\bibinfo {volume} {13}},\ \bibinfo {pages} {3389} (\bibinfo {year}
  {2001})}\BibitemShut {NoStop}%
\bibitem [{\citenamefont {Yamamoto}\ \emph {et~al.}(2006)\citenamefont
  {Yamamoto}, \citenamefont {Stopa}, \citenamefont {Tokura}, \citenamefont
  {Hirayama},\ and\ \citenamefont {Tarucha}}]{Yamamoto2006}%
  \BibitemOpen
  \bibfield  {author} {\bibinfo {author} {\bibfnamefont {M.}~\bibnamefont
  {Yamamoto}}, \bibinfo {author} {\bibfnamefont {M.}~\bibnamefont {Stopa}},
  \bibinfo {author} {\bibfnamefont {Y.}~\bibnamefont {Tokura}}, \bibinfo
  {author} {\bibfnamefont {Y.}~\bibnamefont {Hirayama}}, \ and\ \bibinfo
  {author} {\bibfnamefont {S.}~\bibnamefont {Tarucha}},\ }\href {\doibase
  10.1126/science.1126601} {\bibfield  {journal} {\bibinfo  {journal}
  {Science}\ }\textbf {\bibinfo {volume} {313}},\ \bibinfo {pages} {204}
  (\bibinfo {year} {2006})}\BibitemShut {NoStop}%
\bibitem [{\citenamefont {Laroche}\ \emph {et~al.}(2011)\citenamefont
  {Laroche}, \citenamefont {Gervais}, \citenamefont {Lilly},\ and\
  \citenamefont {Reno}}]{Laroche2011}%
  \BibitemOpen
  \bibfield  {author} {\bibinfo {author} {\bibfnamefont {D.}~\bibnamefont
  {Laroche}}, \bibinfo {author} {\bibfnamefont {G.}~\bibnamefont {Gervais}},
  \bibinfo {author} {\bibfnamefont {M.~P.}\ \bibnamefont {Lilly}}, \ and\
  \bibinfo {author} {\bibfnamefont {J.~L.}\ \bibnamefont {Reno}},\ }\href
  {\doibase 10.1038/nnano.2011.182} {\bibfield  {journal} {\bibinfo  {journal}
  {Nature Nanotech.}\ }\textbf {\bibinfo {volume} {6}},\ \bibinfo {pages} {793}
  (\bibinfo {year} {2011})}\BibitemShut {NoStop}%
\bibitem [{\citenamefont {Laroche}\ \emph {et~al.}(2014)\citenamefont
  {Laroche}, \citenamefont {Gervais}, \citenamefont {Lilly},\ and\
  \citenamefont {Reno}}]{Laroche2014}%
  \BibitemOpen
  \bibfield  {author} {\bibinfo {author} {\bibfnamefont {D.}~\bibnamefont
  {Laroche}}, \bibinfo {author} {\bibfnamefont {G.}~\bibnamefont {Gervais}},
  \bibinfo {author} {\bibfnamefont {M.~P.}\ \bibnamefont {Lilly}}, \ and\
  \bibinfo {author} {\bibfnamefont {J.~L.}\ \bibnamefont {Reno}},\ }\href@noop
  {} {\bibfield  {journal} {\bibinfo  {journal} {Science}\ }\textbf {\bibinfo
  {volume} {343}},\ \bibinfo {pages} {631} (\bibinfo {year}
  {2014})}\BibitemShut {NoStop}%
\bibitem [{\citenamefont {Shinkai}\ \emph {et~al.}(2009)\citenamefont
  {Shinkai}, \citenamefont {Hayashi}, \citenamefont {Ota}, \citenamefont
  {Muraki},\ and\ \citenamefont {Fujisawa}}]{Shinkai2009}%
  \BibitemOpen
  \bibfield  {author} {\bibinfo {author} {\bibfnamefont {G.}~\bibnamefont
  {Shinkai}}, \bibinfo {author} {\bibfnamefont {T.}~\bibnamefont {Hayashi}},
  \bibinfo {author} {\bibfnamefont {T.}~\bibnamefont {Ota}}, \bibinfo {author}
  {\bibfnamefont {K.}~\bibnamefont {Muraki}}, \ and\ \bibinfo {author}
  {\bibfnamefont {T.}~\bibnamefont {Fujisawa}},\ }\href {\doibase
  10.1143/APEX.2.081101} {\bibfield  {journal} {\bibinfo  {journal} {Appl.
  Phys. Express}\ }\textbf {\bibinfo {volume} {2}},\ \bibinfo {pages} {081101}
  (\bibinfo {year} {2009})}\BibitemShut {NoStop}%
\bibitem [{\citenamefont {Nandi}\ \emph {et~al.}(2012)\citenamefont {Nandi},
  \citenamefont {Finck}, \citenamefont {Eisenstein}, \citenamefont {Pfeiffer},\
  and\ \citenamefont {West}}]{Nandi2012}%
  \BibitemOpen
  \bibfield  {author} {\bibinfo {author} {\bibfnamefont {D.}~\bibnamefont
  {Nandi}}, \bibinfo {author} {\bibfnamefont {A.~D.~K.}\ \bibnamefont {Finck}},
  \bibinfo {author} {\bibfnamefont {J.~P.}\ \bibnamefont {Eisenstein}},
  \bibinfo {author} {\bibfnamefont {L.~N.}\ \bibnamefont {Pfeiffer}}, \ and\
  \bibinfo {author} {\bibfnamefont {K.~W.}\ \bibnamefont {West}},\ }\href
  {\doibase 10.1038/nature11302} {\bibfield  {journal} {\bibinfo  {journal}
  {Nature}\ }\textbf {\bibinfo {volume} {488}},\ \bibinfo {pages} {481}
  (\bibinfo {year} {2012})}\BibitemShut {NoStop}%
\bibitem [{\citenamefont {Hartmann}\ \emph {et~al.}(2015)\citenamefont
  {Hartmann}, \citenamefont {Pfeffer}, \citenamefont {H\"ofling}, \citenamefont
  {Kamp},\ and\ \citenamefont {Worschech}}]{Hartmann2015}%
  \BibitemOpen
  \bibfield  {author} {\bibinfo {author} {\bibfnamefont {F.}~\bibnamefont
  {Hartmann}}, \bibinfo {author} {\bibfnamefont {P.}~\bibnamefont {Pfeffer}},
  \bibinfo {author} {\bibfnamefont {S.}~\bibnamefont {H\"ofling}}, \bibinfo
  {author} {\bibfnamefont {M.}~\bibnamefont {Kamp}}, \ and\ \bibinfo {author}
  {\bibfnamefont {L.}~\bibnamefont {Worschech}},\ }\href {\doibase
  10.1103/PhysRevLett.114.146805} {\bibfield  {journal} {\bibinfo  {journal}
  {Phys. Rev. Lett.}\ }\textbf {\bibinfo {volume} {114}},\ \bibinfo {pages}
  {146805} (\bibinfo {year} {2015})}\BibitemShut {NoStop}%
\bibitem [{\citenamefont {Thierschmann}\ \emph {et~al.}(2015)\citenamefont
  {Thierschmann}, \citenamefont {S{\'{a}}nchez}, \citenamefont {Sothmann},
  \citenamefont {Arnold}, \citenamefont {Heyn}, \citenamefont {Hansen},
  \citenamefont {Buhmann},\ and\ \citenamefont {Molenkamp}}]{Thierschmann2015}%
  \BibitemOpen
  \bibfield  {author} {\bibinfo {author} {\bibfnamefont {H.}~\bibnamefont
  {Thierschmann}}, \bibinfo {author} {\bibfnamefont {R.}~\bibnamefont
  {S{\'{a}}nchez}}, \bibinfo {author} {\bibfnamefont {B.}~\bibnamefont
  {Sothmann}}, \bibinfo {author} {\bibfnamefont {F.}~\bibnamefont {Arnold}},
  \bibinfo {author} {\bibfnamefont {C.}~\bibnamefont {Heyn}}, \bibinfo {author}
  {\bibfnamefont {W.}~\bibnamefont {Hansen}}, \bibinfo {author} {\bibfnamefont
  {H.}~\bibnamefont {Buhmann}}, \ and\ \bibinfo {author} {\bibfnamefont
  {L.~W.}\ \bibnamefont {Molenkamp}},\ }\href
  {http://dx.doi.org/10.1038/nnano.2015.176} {\bibfield  {journal} {\bibinfo
  {journal} {Nature Nanotech.}\ }\textbf {\bibinfo {volume} {10}},\ \bibinfo
  {pages} {854} (\bibinfo {year} {2015})}\BibitemShut {NoStop}%
\bibitem [{\citenamefont {Linke}\ \emph {et~al.}(1999)\citenamefont {Linke},
  \citenamefont {Humphrey}, \citenamefont {L{\"{o}}fgren}, \citenamefont
  {Sushkov}, \citenamefont {Newbury}, \citenamefont {Taylor},\ and\
  \citenamefont {Omling}}]{Linke1999}%
  \BibitemOpen
  \bibfield  {author} {\bibinfo {author} {\bibfnamefont {H.}~\bibnamefont
  {Linke}}, \bibinfo {author} {\bibfnamefont {T.~E.}\ \bibnamefont {Humphrey}},
  \bibinfo {author} {\bibfnamefont {A.}~\bibnamefont {L{\"{o}}fgren}}, \bibinfo
  {author} {\bibfnamefont {A.~O.}\ \bibnamefont {Sushkov}}, \bibinfo {author}
  {\bibfnamefont {R.}~\bibnamefont {Newbury}}, \bibinfo {author} {\bibfnamefont
  {R.~P.}\ \bibnamefont {Taylor}}, \ and\ \bibinfo {author} {\bibfnamefont
  {P.}~\bibnamefont {Omling}},\ }\href {\doibase 10.1126/science.286.5448.2314}
  {\bibfield  {journal} {\bibinfo  {journal} {Science}\ }\textbf {\bibinfo
  {volume} {286}},\ \bibinfo {pages} {2314} (\bibinfo {year}
  {1999})}\BibitemShut {NoStop}%
\bibitem [{\citenamefont {Onac}\ \emph {et~al.}(2006)\citenamefont {Onac},
  \citenamefont {Balestro}, \citenamefont {{Willems van Beveren}},
  \citenamefont {Hartmann}, \citenamefont {Nazarov},\ and\ \citenamefont
  {Kouwenhoven}}]{Onac2006}%
  \BibitemOpen
  \bibfield  {author} {\bibinfo {author} {\bibfnamefont {E.}~\bibnamefont
  {Onac}}, \bibinfo {author} {\bibfnamefont {F.}~\bibnamefont {Balestro}},
  \bibinfo {author} {\bibfnamefont {L.~H.}\ \bibnamefont {{Willems van
  Beveren}}}, \bibinfo {author} {\bibfnamefont {U.}~\bibnamefont {Hartmann}},
  \bibinfo {author} {\bibfnamefont {Y.~V.}\ \bibnamefont {Nazarov}}, \ and\
  \bibinfo {author} {\bibfnamefont {L.~P.}\ \bibnamefont {Kouwenhoven}},\
  }\href {http://dx.doi.org/10.1103/PhysRevLett.96.176601} {\bibfield
  {journal} {\bibinfo  {journal} {Phys. Rev. Lett.}\ }\textbf {\bibinfo
  {volume} {96}},\ \bibinfo {pages} {176601} (\bibinfo {year}
  {2006})}\BibitemShut {NoStop}%
\bibitem [{\citenamefont {Khrapai}\ \emph {et~al.}(2006)\citenamefont
  {Khrapai}, \citenamefont {Ludwig}, \citenamefont {Kotthaus}, \citenamefont
  {Tranitz},\ and\ \citenamefont {Wegscheider}}]{Khrapai2006}%
  \BibitemOpen
  \bibfield  {author} {\bibinfo {author} {\bibfnamefont {V.~S.}\ \bibnamefont
  {Khrapai}}, \bibinfo {author} {\bibfnamefont {S.}~\bibnamefont {Ludwig}},
  \bibinfo {author} {\bibfnamefont {J.~P.}\ \bibnamefont {Kotthaus}}, \bibinfo
  {author} {\bibfnamefont {H.~P.}\ \bibnamefont {Tranitz}}, \ and\ \bibinfo
  {author} {\bibfnamefont {W.}~\bibnamefont {Wegscheider}},\ }\href {\doibase
  10.1103/PhysRevLett.97.176803} {\bibfield  {journal} {\bibinfo  {journal}
  {Phys. Rev. Lett.}\ }\textbf {\bibinfo {volume} {97}},\ \bibinfo {pages}
  {176803} (\bibinfo {year} {2006})}\BibitemShut {NoStop}%
\bibitem [{\citenamefont {Roche}\ \emph {et~al.}(2015)\citenamefont {Roche},
  \citenamefont {Roulleau}, \citenamefont {Jullien}, \citenamefont {Jompol},
  \citenamefont {Farrer}, \citenamefont {Ritchie},\ and\ \citenamefont
  {Glattli}}]{Roche2015}%
  \BibitemOpen
  \bibfield  {author} {\bibinfo {author} {\bibfnamefont {B.}~\bibnamefont
  {Roche}}, \bibinfo {author} {\bibfnamefont {P.}~\bibnamefont {Roulleau}},
  \bibinfo {author} {\bibfnamefont {T.}~\bibnamefont {Jullien}}, \bibinfo
  {author} {\bibfnamefont {Y.}~\bibnamefont {Jompol}}, \bibinfo {author}
  {\bibfnamefont {I.}~\bibnamefont {Farrer}}, \bibinfo {author} {\bibfnamefont
  {D.~A.}\ \bibnamefont {Ritchie}}, \ and\ \bibinfo {author} {\bibfnamefont
  {D.~C.}\ \bibnamefont {Glattli}},\ }\href
  {http://dx.doi.org/10.1038/ncomms7738} {\bibfield  {journal} {\bibinfo
  {journal} {Nat. Commun.}\ }\textbf {\bibinfo {volume} {6}} (\bibinfo {year}
  {2015})}\BibitemShut {NoStop}%
\bibitem [{\citenamefont {Costache}\ and\ \citenamefont
  {Valenzuela}(2010)}]{Costache2010}%
  \BibitemOpen
  \bibfield  {author} {\bibinfo {author} {\bibfnamefont {M.~V.}\ \bibnamefont
  {Costache}}\ and\ \bibinfo {author} {\bibfnamefont {S.~O.}\ \bibnamefont
  {Valenzuela}},\ }\href {\doibase 10.1126/science.1196228} {\bibfield
  {journal} {\bibinfo  {journal} {Science}\ }\textbf {\bibinfo {volume}
  {330}},\ \bibinfo {pages} {1645} (\bibinfo {year} {2010})}\BibitemShut
  {NoStop}%
\bibitem [{\citenamefont {Rousselet}\ \emph {et~al.}(1994)\citenamefont
  {Rousselet}, \citenamefont {Salome}, \citenamefont {Ajdari},\ and\
  \citenamefont {Prostt}}]{Rousselet1994}%
  \BibitemOpen
  \bibfield  {author} {\bibinfo {author} {\bibfnamefont {J.}~\bibnamefont
  {Rousselet}}, \bibinfo {author} {\bibfnamefont {L.}~\bibnamefont {Salome}},
  \bibinfo {author} {\bibfnamefont {A.}~\bibnamefont {Ajdari}}, \ and\ \bibinfo
  {author} {\bibfnamefont {J.}~\bibnamefont {Prostt}},\ }\href {\doibase
  10.1038/370446a0} {\bibfield  {journal} {\bibinfo  {journal} {Nature}\
  }\textbf {\bibinfo {volume} {370}},\ \bibinfo {pages} {446} (\bibinfo {year}
  {1994})}\BibitemShut {NoStop}%
\bibitem [{\citenamefont {Bischoff}\ \emph {et~al.}(2015)\citenamefont
  {Bischoff}, \citenamefont {Eich}, \citenamefont {Zilberberg}, \citenamefont
  {R{\"{o}}ssler}, \citenamefont {Ihn},\ and\ \citenamefont
  {Ensslin}}]{Bischoff2015}%
  \BibitemOpen
  \bibfield  {author} {\bibinfo {author} {\bibfnamefont {D.}~\bibnamefont
  {Bischoff}}, \bibinfo {author} {\bibfnamefont {M.}~\bibnamefont {Eich}},
  \bibinfo {author} {\bibfnamefont {O.}~\bibnamefont {Zilberberg}}, \bibinfo
  {author} {\bibfnamefont {C.}~\bibnamefont {R{\"{o}}ssler}}, \bibinfo {author}
  {\bibfnamefont {T.}~\bibnamefont {Ihn}}, \ and\ \bibinfo {author}
  {\bibfnamefont {K.}~\bibnamefont {Ensslin}},\ }\href {\doibase
  10.1021/acs.nanolett.5b02167} {\bibfield  {journal} {\bibinfo  {journal}
  {Nano Lett.}\ }\textbf {\bibinfo {volume} {15}},\ \bibinfo {pages} {6003}
  (\bibinfo {year} {2015})}\BibitemShut {NoStop}%
\bibitem [{\citenamefont {Kaasbjerg}\ and\ \citenamefont
  {Jauho}(2016)}]{kaa16}%
  \BibitemOpen
  \bibfield  {author} {\bibinfo {author} {\bibfnamefont {K.}~\bibnamefont
  {Kaasbjerg}}\ and\ \bibinfo {author} {\bibfnamefont {A.-P.}\ \bibnamefont
  {Jauho}},\ }\href@noop {} {\bibfield  {journal} {\bibinfo  {journal} {Phys.
  Rev. Lett.}\ }\textbf {\bibinfo {volume} {116}},\ \bibinfo {pages} {196801}
  (\bibinfo {year} {2016})}\BibitemShut {NoStop}%
\bibitem [{\citenamefont {Amasha}\ \emph {et~al.}(2013)\citenamefont {Amasha},
  \citenamefont {Keller}, \citenamefont {Rau}, \citenamefont {Carmi},
  \citenamefont {Katine}, \citenamefont {Shtrikman}, \citenamefont {Oreg},\
  and\ \citenamefont {Goldhaber-Gordon}}]{Amasha2013:Pseudospin}%
  \BibitemOpen
  \bibfield  {author} {\bibinfo {author} {\bibfnamefont {S.}~\bibnamefont
  {Amasha}}, \bibinfo {author} {\bibfnamefont {A.~J.}\ \bibnamefont {Keller}},
  \bibinfo {author} {\bibfnamefont {I.~G.}\ \bibnamefont {Rau}}, \bibinfo
  {author} {\bibfnamefont {A.}~\bibnamefont {Carmi}}, \bibinfo {author}
  {\bibfnamefont {J.~A.}\ \bibnamefont {Katine}}, \bibinfo {author}
  {\bibfnamefont {H.}~\bibnamefont {Shtrikman}}, \bibinfo {author}
  {\bibfnamefont {Y.}~\bibnamefont {Oreg}}, \ and\ \bibinfo {author}
  {\bibfnamefont {D.}~\bibnamefont {Goldhaber-Gordon}},\ }\href {\doibase
  10.1103/PhysRevLett.110.046604} {\bibfield  {journal} {\bibinfo  {journal}
  {Phys. Rev. Lett.}\ }\textbf {\bibinfo {volume} {110}},\ \bibinfo {pages}
  {046604} (\bibinfo {year} {2013})}\BibitemShut {NoStop}%
\bibitem [{\citenamefont {Keller}\ \emph {et~al.}(2014)\citenamefont {Keller},
  \citenamefont {Amasha}, \citenamefont {Weymann}, \citenamefont {Moca},
  \citenamefont {Rau}, \citenamefont {Katine}, \citenamefont {Shtrikman},
  \citenamefont {Z\'arand},\ and\ \citenamefont
  {Goldhaber-Gordon}}]{Keller2014:EmergentSU4}%
  \BibitemOpen
  \bibfield  {author} {\bibinfo {author} {\bibfnamefont {A.~J.}\ \bibnamefont
  {Keller}}, \bibinfo {author} {\bibfnamefont {S.}~\bibnamefont {Amasha}},
  \bibinfo {author} {\bibfnamefont {I.}~\bibnamefont {Weymann}}, \bibinfo
  {author} {\bibfnamefont {C.~P.}\ \bibnamefont {Moca}}, \bibinfo {author}
  {\bibfnamefont {I.~G.}\ \bibnamefont {Rau}}, \bibinfo {author} {\bibfnamefont
  {J.~A.}\ \bibnamefont {Katine}}, \bibinfo {author} {\bibfnamefont
  {H.}~\bibnamefont {Shtrikman}}, \bibinfo {author} {\bibfnamefont
  {G.}~\bibnamefont {Z\'arand}}, \ and\ \bibinfo {author} {\bibfnamefont
  {D.}~\bibnamefont {Goldhaber-Gordon}},\ }\href {\doibase 10.1038/nphys2844}
  {\bibfield  {journal} {\bibinfo  {journal} {Nature Phys.}\ }\textbf {\bibinfo
  {volume} {10}},\ \bibinfo {pages} {145} (\bibinfo {year} {2014})}\BibitemShut
  {NoStop}%
\bibitem [{bia()}]{biasnote}%
  \BibitemOpen
  \href@noop {} {}\bibinfo {note} {Note that we apply $V_{S1}$ while
  compensating with gate voltages $V_{P1},V_{P2}$ to avoid changing either of
  the dot levels, though the ac-part of the bias may still act as an
  ac-gate.}\BibitemShut {Stop}%
\bibitem [{SI()}]{SI}%
  \BibitemOpen
  \href@noop {} {}\bibinfo {note} {See Supplemental Material for additional
  data and further theoretical details, which includes
  Ref.~\cite{vanderwiel}.}\BibitemShut {Stop}%
\bibitem [{\citenamefont {Bruus}\ and\ \citenamefont
  {Flensberg}(2004)}]{bruus}%
  \BibitemOpen
  \bibfield  {author} {\bibinfo {author} {\bibfnamefont {H.}~\bibnamefont
  {Bruus}}\ and\ \bibinfo {author} {\bibfnamefont {K.}~\bibnamefont
  {Flensberg}},\ }\href@noop {} {\emph {\bibinfo {title} {Many-Body Quantum
  Theory in Condensed Matter Physics}}}\ (\bibinfo  {publisher} {Oxford
  University Press, Oxford, UK},\ \bibinfo {year} {2004})\BibitemShut {NoStop}%
\bibitem [{\citenamefont {Averin}(1994)}]{averin}%
  \BibitemOpen
  \bibfield  {author} {\bibinfo {author} {\bibfnamefont {D.}~\bibnamefont
  {Averin}},\ }\href@noop {} {\bibfield  {journal} {\bibinfo  {journal}
  {Physica B}\ }\textbf {\bibinfo {volume} {194-196}},\ \bibinfo {pages} {979}
  (\bibinfo {year} {1994})}\BibitemShut {NoStop}%
\bibitem [{\citenamefont {Turek}\ and\ \citenamefont {Matveev}(2002)}]{turek}%
  \BibitemOpen
  \bibfield  {author} {\bibinfo {author} {\bibfnamefont {M.}~\bibnamefont
  {Turek}}\ and\ \bibinfo {author} {\bibfnamefont {K.~A.}\ \bibnamefont
  {Matveev}},\ }\href@noop {} {\bibfield  {journal} {\bibinfo  {journal} {Phys.
  Rev. B}\ }\textbf {\bibinfo {volume} {65}},\ \bibinfo {pages} {115332}
  (\bibinfo {year} {2002})}\BibitemShut {NoStop}%
\bibitem [{not()}]{note}%
  \BibitemOpen
  \href@noop {} {}\bibinfo {note} {The only different parameter is temperature.
  We take $k_B T=5\Gamma$ to ensure the validity of our theory. Note that the
  experiments show a weak dependence of $I_{\rm drag}$ with $k_B T$
  [Fig.~\ref{fig:deps}(a) and~(b)].}\BibitemShut {Stop}%
\bibitem [{\citenamefont {Van~der Wiel}\ \emph {et~al.}(2002)\citenamefont
  {Van~der Wiel}, \citenamefont {De~Franceschi}, \citenamefont {Elzerman},
  \citenamefont {Fujisawa}, \citenamefont {Tarucha},\ and\ \citenamefont
  {Kouwenhoven}}]{vanderwiel}%
  \BibitemOpen
  \bibfield  {author} {\bibinfo {author} {\bibfnamefont {W.~G.}\ \bibnamefont
  {Van~der Wiel}}, \bibinfo {author} {\bibfnamefont {S.}~\bibnamefont
  {De~Franceschi}}, \bibinfo {author} {\bibfnamefont {J.~M.}\ \bibnamefont
  {Elzerman}}, \bibinfo {author} {\bibfnamefont {T.}~\bibnamefont {Fujisawa}},
  \bibinfo {author} {\bibfnamefont {S.}~\bibnamefont {Tarucha}}, \ and\
  \bibinfo {author} {\bibfnamefont {L.~P.}\ \bibnamefont {Kouwenhoven}},\
  }\href {\doibase 10.1103/RevModPhys.75.1} {\bibfield  {journal} {\bibinfo
  {journal} {Rev. Mod. Phys.}\ }\textbf {\bibinfo {volume} {75}},\ \bibinfo
  {pages} {1} (\bibinfo {year} {2002})}\BibitemShut {NoStop}%
\end{thebibliography}%


\newcommand{\noopsort}[1]{} \newcommand{\printfirst}[2]{#1}
  \newcommand{\singleletter}[1]{#1} \newcommand{\switchargs}[2]{#2#1}
\begin{thebibliography}{3}%
\makeatletter
\providecommand \@ifxundefined [1]{%
 \@ifx{#1\undefined}
}%
\providecommand \@ifnum [1]{%
 \ifnum #1\expandafter \@firstoftwo
 \else \expandafter \@secondoftwo
 \fi
}%
\providecommand \@ifx [1]{%
 \ifx #1\expandafter \@firstoftwo
 \else \expandafter \@secondoftwo
 \fi
}%
\providecommand \natexlab [1]{#1}%
\providecommand \enquote  [1]{``#1''}%
\providecommand \bibnamefont  [1]{#1}%
\providecommand \bibfnamefont [1]{#1}%
\providecommand \citenamefont [1]{#1}%
\providecommand \href@noop [0]{\@secondoftwo}%
\providecommand \href [0]{\begingroup \@sanitize@url \@href}%
\providecommand \@href[1]{\@@startlink{#1}\@@href}%
\providecommand \@@href[1]{\endgroup#1\@@endlink}%
\providecommand \@sanitize@url [0]{\catcode `\\12\catcode `\$12\catcode
  `\&12\catcode `\#12\catcode `\^12\catcode `\_12\catcode `\%12\relax}%
\providecommand \@@startlink[1]{}%
\providecommand \@@endlink[0]{}%
\providecommand \url  [0]{\begingroup\@sanitize@url \@url }%
\providecommand \@url [1]{\endgroup\@href {#1}{\urlprefix }}%
\providecommand \urlprefix  [0]{URL }%
\providecommand \Eprint [0]{\href }%
\providecommand \doibase [0]{http://dx.doi.org/}%
\providecommand \selectlanguage [0]{\@gobble}%
\providecommand \bibinfo  [0]{\@secondoftwo}%
\providecommand \bibfield  [0]{\@secondoftwo}%
\providecommand \translation [1]{[#1]}%
\providecommand \BibitemOpen [0]{}%
\providecommand \bibitemStop [0]{}%
\providecommand \bibitemNoStop [0]{.\EOS\space}%
\providecommand \EOS [0]{\spacefactor3000\relax}%
\providecommand \BibitemShut  [1]{\csname bibitem#1\endcsname}%
\let\auto@bib@innerbib\@empty
\bibitem [{\citenamefont {Van~der Wiel}\ \emph {et~al.}(2002)\citenamefont
  {Van~der Wiel}, \citenamefont {De~Franceschi}, \citenamefont {Elzerman},
  \citenamefont {Fujisawa}, \citenamefont {Tarucha},\ and\ \citenamefont
  {Kouwenhoven}}]{vanderwiel}%
  \BibitemOpen
  \bibfield  {author} {\bibinfo {author} {\bibfnamefont {W.~G.}\ \bibnamefont
  {Van~der Wiel}}, \bibinfo {author} {\bibfnamefont {S.}~\bibnamefont
  {De~Franceschi}}, \bibinfo {author} {\bibfnamefont {J.~M.}\ \bibnamefont
  {Elzerman}}, \bibinfo {author} {\bibfnamefont {T.}~\bibnamefont {Fujisawa}},
  \bibinfo {author} {\bibfnamefont {S.}~\bibnamefont {Tarucha}}, \ and\
  \bibinfo {author} {\bibfnamefont {L.~P.}\ \bibnamefont {Kouwenhoven}},\
  }\href {\doibase 10.1103/RevModPhys.75.1} {\bibfield  {journal} {\bibinfo
  {journal} {Rev. Mod. Phys.}\ }\textbf {\bibinfo {volume} {75}},\ \bibinfo
  {pages} {1} (\bibinfo {year} {2002})}\BibitemShut {NoStop}%
\bibitem [{\citenamefont {Amasha}\ \emph {et~al.}(2013)\citenamefont {Amasha},
  \citenamefont {Keller}, \citenamefont {Rau}, \citenamefont {Carmi},
  \citenamefont {Katine}, \citenamefont {Shtrikman}, \citenamefont {Oreg},\
  and\ \citenamefont {Goldhaber-Gordon}}]{Amasha2013:Pseudospin}%
  \BibitemOpen
  \bibfield  {author} {\bibinfo {author} {\bibfnamefont {S.}~\bibnamefont
  {Amasha}}, \bibinfo {author} {\bibfnamefont {A.~J.}\ \bibnamefont {Keller}},
  \bibinfo {author} {\bibfnamefont {I.~G.}\ \bibnamefont {Rau}}, \bibinfo
  {author} {\bibfnamefont {A.}~\bibnamefont {Carmi}}, \bibinfo {author}
  {\bibfnamefont {J.~A.}\ \bibnamefont {Katine}}, \bibinfo {author}
  {\bibfnamefont {H.}~\bibnamefont {Shtrikman}}, \bibinfo {author}
  {\bibfnamefont {Y.}~\bibnamefont {Oreg}}, \ and\ \bibinfo {author}
  {\bibfnamefont {D.}~\bibnamefont {Goldhaber-Gordon}},\ }\href {\doibase
  10.1103/PhysRevLett.110.046604} {\bibfield  {journal} {\bibinfo  {journal}
  {Phys. Rev. Lett.}\ }\textbf {\bibinfo {volume} {110}},\ \bibinfo {pages}
  {046604} (\bibinfo {year} {2013})}\BibitemShut {NoStop}%
\bibitem [{\citenamefont {Kaasbjerg}\ and\ \citenamefont
  {Jauho}(2016)}]{kaa16}%
  \BibitemOpen
  \bibfield  {author} {\bibinfo {author} {\bibfnamefont {K.}~\bibnamefont
  {Kaasbjerg}}\ and\ \bibinfo {author} {\bibfnamefont {A.-P.}\ \bibnamefont
  {Jauho}},\ }\href@noop {} {\bibfield  {journal} {\bibinfo  {journal} {Phys.
  Rev. Lett.}\ }\textbf {\bibinfo {volume} {116}},\ \bibinfo {pages} {196801}
  (\bibinfo {year} {2016})}\BibitemShut {NoStop}%
\end{thebibliography}%
\end{document}


\title{Supplemental information: Cotunneling drag effect in \protect{C}oulomb-coupled quantum dots}
\author{A. J. Keller}
	\altaffiliation{Present address: Institute for Quantum Information and Matter, California Institute of Technology, Pasadena, California 91125, USA}
	\affiliation{Department of Physics, Stanford University, Stanford, California 94305, USA}

\author{J. S. Lim}
	\affiliation{School of Physics, Korea Institute for Advanced Study, Seoul 130-722, Korea}

\author{David S\'anchez}
	\affiliation{IFISC (UIB-CSIC), E-07122 Palma de Mallorca, Spain}
	
\author{Rosa L\'opez}
	\affiliation{IFISC (UIB-CSIC), E-07122 Palma de Mallorca, Spain}

\author{S. Amasha}
	\altaffiliation{Present address: MIT Lincoln Laboratory, Lexington, Massachusetts 02420, USA}
	\affiliation{Department of Physics, Stanford University, Stanford, California 94305, USA}

\author{J. A. Katine}
	\affiliation{HGST, San Jose, CA 95135, USA}

\author{Hadas Shtrikman} 
	\affiliation{Department of Condensed Matter Physics, Weizmann Institute of Science, Rehovot 96100, Israel}

\author{D. Goldhaber-Gordon}
	\email{goldhaber-gordon@stanford.edu}
	\affiliation{Department of Physics, Stanford University, Stanford, California 94305, USA}

\maketitle

\onecolumngrid
\section{Capacitances}

To extract the capacitances in our device, we begin by assuming the usual electrostatic model for a double quantum dot (DQD)~\cite{vanderwiel}, which has a capacitor between each dot and its source and drain leads ($C_{Si}$ and $C_{Di}$, for $i \in \{1,2\}$), each dot and its gate ($C_{Pi}$), and a capacitor between the dots ($C_m$). We define $C_i$ to be the total capacitance of dot $i$, e.g. $C_1 = C_{P1} + C_{S1} + C_{m} +$ capacitances to ground. 

The electrostatic energy of the system $\vec{Q} \cdot \vec{V}/2$ is given by:
%
\beq
\begin{aligned}
\widetilde{U}(N_1,N_2) &= \frac{1}{2} U_1 N_1^2 + \frac{1}{2} U_2 N_2^2 + U N_1 N_2 -\frac{1}{2|e|} \sum\limits_{i \in \{1,2\}} U_i \left\{ C_{Pi} V_{Pi} N_i + C_{Si} V_{Si} N_i \right\} \\
&-\frac{1}{2|e|} U \left\{ C_{P1} V_{P1} N_2 + C_{P2} V_{P2} N_1 + C_{S1} V_{S1} N_2 + C_{S2} V_{S2} N_1 \right\}
\end{aligned}
\label{eqn:U}
\eeq
%
where we let $Q_i = -|e| N_i$. The addition energies of the system are given by:
\beq
U_1 = e^2 \frac{C_2}{C_1 C_2 - C_m^2} = \frac{e^2}{C_1} X
\eeq
\beq
U_2 = e^2 \frac{C_1}{C_1 C_2 - C_m^2} = \frac{e^2}{C_2} X
\eeq
\beq
U = e^2 \frac{C_m}{C_1 C_2 - C_m^2} = \frac{e^2}{C_m} (X-1)
\eeq
where
\beq
X = \frac{1}{\left(1-\frac{C_m^2}{C_1 C_2}\right)}
\eeq
It follows that $C_m/C_2 = U/U_1$ and $C_m/C_1 = U/U_2$, so we may also write
\beq
X = \frac{1}{\left(1-\frac{U^2}{U_1U_2}\right)}
\eeq
The addition energies can may be extracted simply from transport measurements and we use the values reported in our previous work~\cite{Amasha2013:Pseudospin}. We obtain $X = 1.006$ in terms of the addition energies $U_1 = 1.2$~meV, $U_2 = 1.5$~meV, and $U = 0.1$~meV. We then obtain $C_1 = 130$~aF, $C_2 = 107$~aF, $C_m = 9.0$~aF.

\begin{figure}
\includegraphics[width=3.37in]{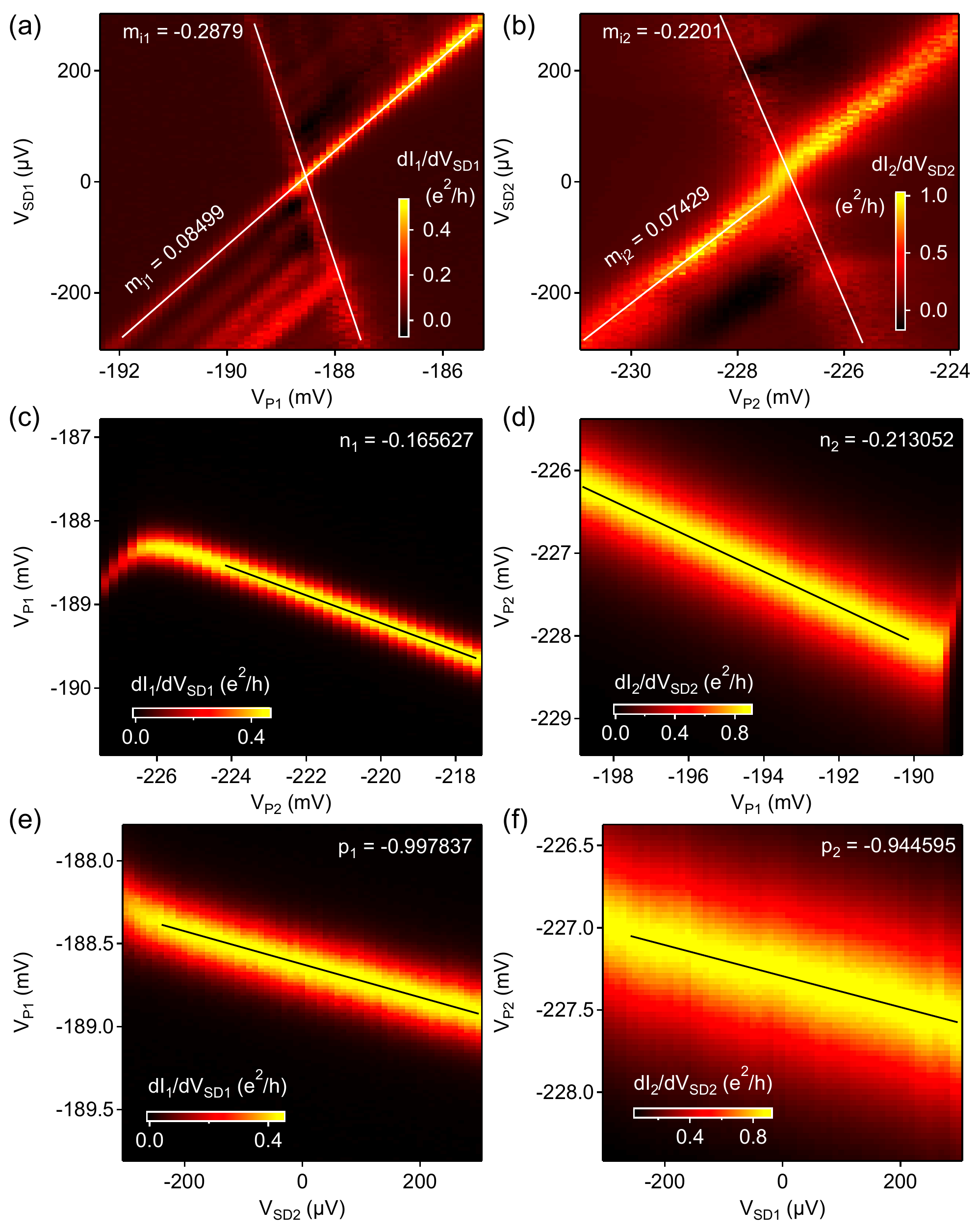}
\caption{Determining capacitances. (a) Source-drain bias spectroscopy for dot 1. $V_{P2} = -232.4$~mV is chosen to be away from the triple points. Slopes $m_{i1},m_{j1}$ are indicated. (b) Source-drain bias spectroscopy for dot 2. $V_{P1} = -193.8$~mV is chosen to be away from the triple points. Slopes $m_{i2},m_{j2}$ are indicated. (c,d) The capacitance of gate P2 to dot 1 (c) and gate P1 to dot 2 (d) may be obtained with these zero source-drain bias measurements. Slopes $n_1, n_2$ are indicated. (e,f) The capacitance of reservoir S2 to dot 1 (e) and S1 to dot 2 (f) may be obtained. Slopes $p_1,p_2$ are indicated. (e) $V_{P2} = -232.4$~mV. (f) $V_{P1} = -193.8$~mV.
\label{fig:calib}}
\end{figure}

The effective dot levels are given by:
\beq
\mu_1 (N_1,N_2) = \widetilde{U}(N_1,N_2) - \widetilde{U}(N_1-1,N_2)
\eeq
\beq
\mu_2 (N_1,N_2) = \widetilde{U}(N_1,N_2) - \widetilde{U}(N_1,N_2-1)
\eeq
The slopes of linear features seen in source-drain bias spectroscopy correspond to conditions on the effective dot levels. The slopes may be related to the capacitances. The slopes $m_{i1}$ and $m_{j1}$ in \subfig{calib}{a} are:
\beq
m_{i1} = -\frac{C_{P1}}{C_{S1}}
\eeq

\beq
m_{j1} = \frac{C_{P1}}{2C_1/X - C_{S1}}
\eeq
These can be used to solve for the remaining capacitances:
\beq
C_{P1} = \frac{2C_1}{X} \frac{m_{i1} m_{j1}}{m_{i1}-m_{j1}}
\eeq

\beq
C_{S1} = -\frac{2C_1}{X} \frac{m_{j1}}{m_{i1}-m_{j1}}
\eeq
The capacitances $C_{P2}$ and $C_{S2}$ are similarly related to the slopes $m_{i2}$ and $m_{j2}$ in \subfig{calib}{b}. We find $C_{P1} = 17.5$~aF, $C_{S1} = 60.9$~aF, $C_{P2} = 11.9$~aF, $C_{S2} = 53.9$~aF. 

At this point we can check for self-consistency using other measurements. \Startsubfig{calib}{c} and \subfig{calib}{d} show part of the charge stability diagram (note the axes are swapped between (c) and (d)). In \subfig{calib}{c}, the peak in $G_1$ corresponds to constant $\mu_1$. The following condition holds:
\beq
U_1 C_{P1} \Delta V_{P1} + U C_{P2} \Delta V_{P2} = 0
\eeq
Therefore we would expect that the measured slope $m = \Delta V_{P1} / \Delta V_{P2} = -0.166$ should equal $-U C_{P2} / U_1 C_{P1} = -0.0567$. The large discrepancy indicates that the typical electrostatic model of a DQD is insufficient to describe our data.

To account for the discrepancy we introduce direct capacitances between reservoir S1 and dot 2 ($C_{S1,2}$), and vice versa. We also introduce capacitances between gate P1 and dot 2 ($C_{P1,2}$), and vice versa, though these will be small. The total capacitances $C_1$ and $C_2$ will now include these extra capacitances. To \refeq{U} we must add corresponding terms:

\beq
\begin{aligned}
& -\frac{1}{2|e|} U_1 \left\{ C_{P2,1} N_1 V_{P2} + C_{S2,1} N_1 V_{S2} \right\} - \frac{1}{2|e|} U_2 \left\{ C_{P1,2} N_2 V_{P1} + C_{S1,2} N_2 V_{S1} \right\} \\
& -\frac{1}{2|e|} U \left\{ C_{P1,2} N_1 V_{P1} + C_{P2,1} N_2 V_{P2} + C_{S1,2} N_1 V_{S1} + C_{S2,1} N_2 V_{S2} \right\}
\end{aligned}
\label{eq:U2}
\eeq

While the definition of $U_1, U_2, U$ are unchanged, we must reinterpret the slopes of the linear features observed in measurements. For \subfig{calib}{a,b} the slopes $m_{jk}, m_{ik}; k \in \{1,2\}$ are given by:

\beq
m_{jk} = \frac{\Delta V_{Sk}}{\Delta V_{Pk}} = \frac{E_{Ck} C_{P1} + U C_{Pk,\bar{k}}}{2e^2-(E_{Ck} C_{Sk} + U C_{Sk,\bar{k}})}
\eeq
\beq
m_{ik} = \frac{\Delta V_{Sk}}{\Delta V_{Pk}} = - \frac{E_{Ck} C_{P1} + U C_{Pk,\bar{k}}}{E_{Ck} C_{Sk} + U C_{Sk,\bar{k}}}
\eeq
where $\bar{k} = 1 (2)$ if $k = 2 (1)$. Additional measurements are needed to solve for all capacitances. The slope of the feature observed in \subfig{calib}{c,d} is:

\beq
n_k = \frac{\Delta V_{Pk}}{\Delta V_{P\bar{k}}} = - \frac{E_{Ck} C_{P\bar{k},k} + U C_{P\bar{k}}}{E_{Ck} C_{Pk} + U C_{Pk,\bar{k}}}
\eeq
where $k = 1$ for (c) and $k = 2$ for (d). We can now solve for four of the capacitances. It may be shown that:

\beq
-n_k \frac{m_{jk} m_{ik}}{m_{ik}-m_{jk}} = \frac{X}{2} \frac{C_{P\bar{k},k}}{C_k} + \frac{X-1}{2} \frac{C_{P\bar{k}}}{C_m}
\label{eqn:peq1}
\eeq
\beq
\frac{m_{jk} m_{ik}}{m_{ik}-m_{jk}} = \frac{X}{2} \frac{C_{Pk}}{C_k} + \frac{X-1}{2} \frac{C_{Pk,\bar{k}}}{C_m}
\label{eqn:peq2}
\eeq
Equations (\ref{eqn:peq1}) and (\ref{eqn:peq2}) may be solved for $C_{Pk}$ and $C_{Pk,\bar{k}}$, yielding $C_{P1} = 17.4$~aF, $C_{P1,2} = 1.37$~aF, $C_{P2} = 11.7$~aF, $C_{P2,1} = 1.92$~aF.

In \subfig{calib}{e,f}, the slopes are given by:

\beq
p_k = \frac{\Delta V_{Pk}}{\Delta V_{SD\bar{k}}} = -\frac{E_{Ck} C_{S\bar{k},k} + U C_{S\bar{k}}}{E_{Ck} C_{Pk} + U C_{Pk,\bar{k}}}
\eeq
It may be shown that:

\beq
-p_k \frac{m_{jk} m_{ik}}{m_{ik}-m_{jk}} = \frac{X}{2} \frac{C_{S\bar{k},k}}{C_k} + \frac{X-1}{2} \frac{C_{S\bar{k}}}{C_m}
\label{eqn:seq1}
\eeq

\beq
- \frac{m_{jk}}{m_{ik}-m_{jk}} = \frac{X}{2} \frac{C_{Sk}}{C_{k}} + \frac{X-1}{2} \frac{C_{Sk,\bar{k}}}{C_m}
\label{eqn:seq2}
\eeq
Solving the system of equations (\ref{eqn:seq1}) and (\ref{eqn:seq2}) yields $C_{S1} = 60.3$~aF, $C_{S2} = 53.0$~aF, $C_{S1,2} = 7.19$~aF, and $C_{S2,1} = 13.1$~aF. We emphasize that capacitances like $C_{S2,1}$ are bigger than or comparable to $C_m$, and cannot be neglected.

The capacitances are summarized in \alltab{caps}. By accounting for all of these capacitances, one can take measurements in such a way as to apply a source-drain bias without gating the dot (changing the effective dot levels), and have independent effective gates for each dot, as done in the manuscript and described in the supplementary information of previous work~\cite{Amasha2013:Pseudospin}.

\begin{table}[h]
\begin{tabular}{cc}
& Value (aF) \\
\hline
$C_1$ & 134 \\
$C_2$ & 107 \\
$C_m$ & 8.97 \\
$C_{P1}$ & 17.4 \\
$C_{P2}$ & 11.7 \\
$C_{P1,2}$ & 1.37 \\
$C_{P2,1}$ & 1.92 \\
$C_{S1}$ & 60.3 \\
$C_{S2}$ & 53.0 \\
$C_{S1,2}$ & 7.19 \\
$C_{S2,1}$ & 13.1 \\
\end{tabular}
\caption{Experimentally derived capacitances. \label{tab:caps}}
\end{table}

\newpage
\section{Limits on interdot tunneling}

\begin{sidewaysfigure}
\includegraphics[width=9in]{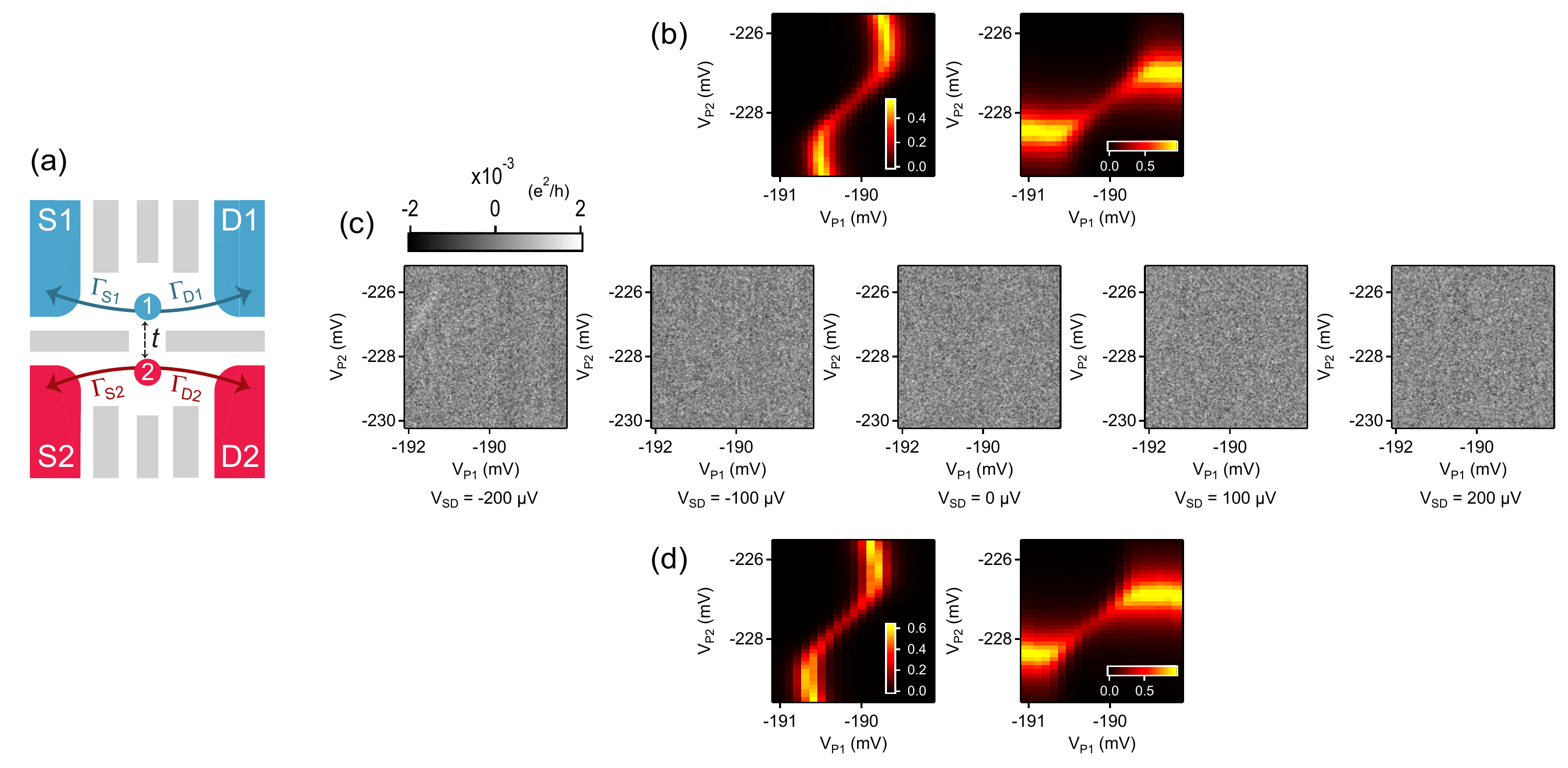}
\caption{Limits on interdot tunneling. (a) Interdot tunneling refers to direct hopping from dot 1 to dot 2 or vice versa. (b) Measured $dI_1/dV_{S1}$ (left) and $dI_2/dV_{S2}$ (right) at zero dc bias near triple points of the DQD. (c) Measured $G_{\rm series} = dI/dV_{SD}$. The bias voltage is applied to both leads of dot 1 and current is measured by one current amplifier connected to both leads of dot 2. Five different dc biases were applied as indicated. No features are observed, except perhaps a faint signal when $V_{SD} = -200$~$\mu$V near the upper left of the plot. (d) Immediately after the measurements of (c), we repeat the measurements of (b) at lower resolution just to ensure that the features have not drifted. $dI_1/dV_{S1}$ is again at left.
\label{fig:tun}}
\end{sidewaysfigure}

The series conductance $G_{\rm series}$ at a triple point of the DQD places a limit on the interdot tunneling energy scale $t$. The series conductance is measured by applying an ac+dc bias voltage to both leads of dot 1 and using one current amplifier attached to both leads of dot 2 to measure current. In the case of zero dc bias~\cite[supp. info]{Amasha2013:Pseudospin}:

\begin{equation}
G_{\textrm{series}} = \frac{64 |t|^2}{3 \Gamma_1 \left( \frac{\Gamma_1+\Gamma_2}{2} \right)} \frac{e^2}{h}
\end{equation}
where $\Gamma_i$ is the total tunnel rate between dot $i$ and its two leads S$,i$ and D$,i$. In \subfig{tun}{a} we present conductance in a region of gate voltage containing the triple points of the DQD. For simplicity, here we are just varying the gate voltages $V_{P1,P2}$, rather than independently controlling $\varepsilon_{1,2}$. We switch measurement configurations in \subfig{tun}{b} and measure $G_{\rm series}$ at several dc biases, including zero bias. In contrast with transport through the individual dots (\subfig{tun}{a}), conductance through the two dots is at our noise floor. Perhaps a very faint signal can be observed for certain gate voltages at $V_{SD} = -200$~$\mu$V, but even there, $G_{\rm series} < 2\times10^{-3}$~$e^2/h$ for all measured dc biases. Given that $\Gamma_1 \sim 15$~$\mu$eV and $\Gamma_2 \sim 47$~$\mu$eV, we find that $|t| < 0.21$~$\mu$eV at zero dc bias. We do not rule out a dependence of $|t|$ on the bias voltage, but at least we do not observe appreciable $G_{\rm series}$ at non-zero bias either. To rule out the possibility that the triple points have drifted out of the measurement window owing to charge instability in the device, we immediately repeat the measurement of (a) in \subfig{tun}{c}, and see that the features have barely moved.

\newpage

\section{Drag at zero field}

\begin{figure}
\includegraphics[width=3.37in]{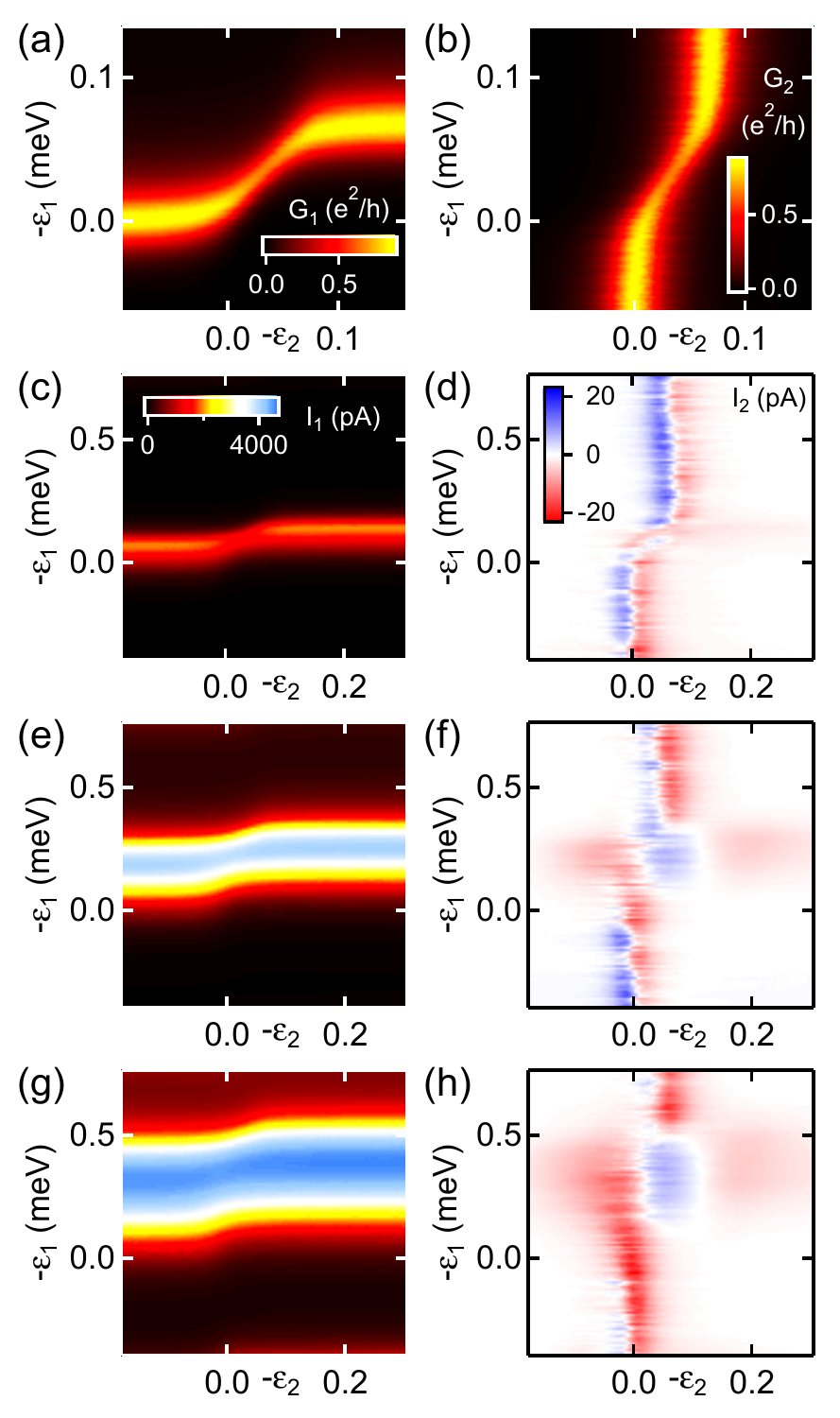}
\caption{Measurements with zero applied magnetic field. (a) Measured $dI_1/dV_{SD1}$ and (b) $dI_2/dV_{SD2}$ at zero dc bias near triple points of the DQD. (c,e,g) Measured $I_1$ and (d,f,h) $I_2$ for: (c,d) $V_{SD1} = 100$~$\mu$V; (e,f) $V_{SD1} = 300$~$\mu$V; (g,h) $V_{SD1} = 500$~$\mu$V. Note that in (c--h) the axis scaling differs from (a,b). No bias is applied across dot 2. In all cases, a drag current can be observed through dot 2 despite no explicit bias being applied to dot 2.
\label{fig:zero}}
\end{figure}

Figures 2--4 show measurements taken in a large Zeeman field and small out-of-plane field. This is a regime where our (spinless) theory applies, and is where we took most drag measurements. For reference, \Startallfig{zero} shows measurements taken in zero applied field. Qualitatively similar features are observed, including a finite drag current $I_2$ many linewidths away from resonance. A notable difference compared to the finite Zeeman field case is that the sign of $I_2$ changes as a function of $-\varepsilon_2$. This should not be surprising: even in the spinless theory we find that depending on tunnel couplings the direction of drag current can reverse in this way. This is true for both our model and the model in \cite{kaa16} (see their Fig. 3b).

\section{Theoretical formalism}

\subsection{Electrostatic model}
\begin{figure}
\includegraphics[width=0.45\textwidth]{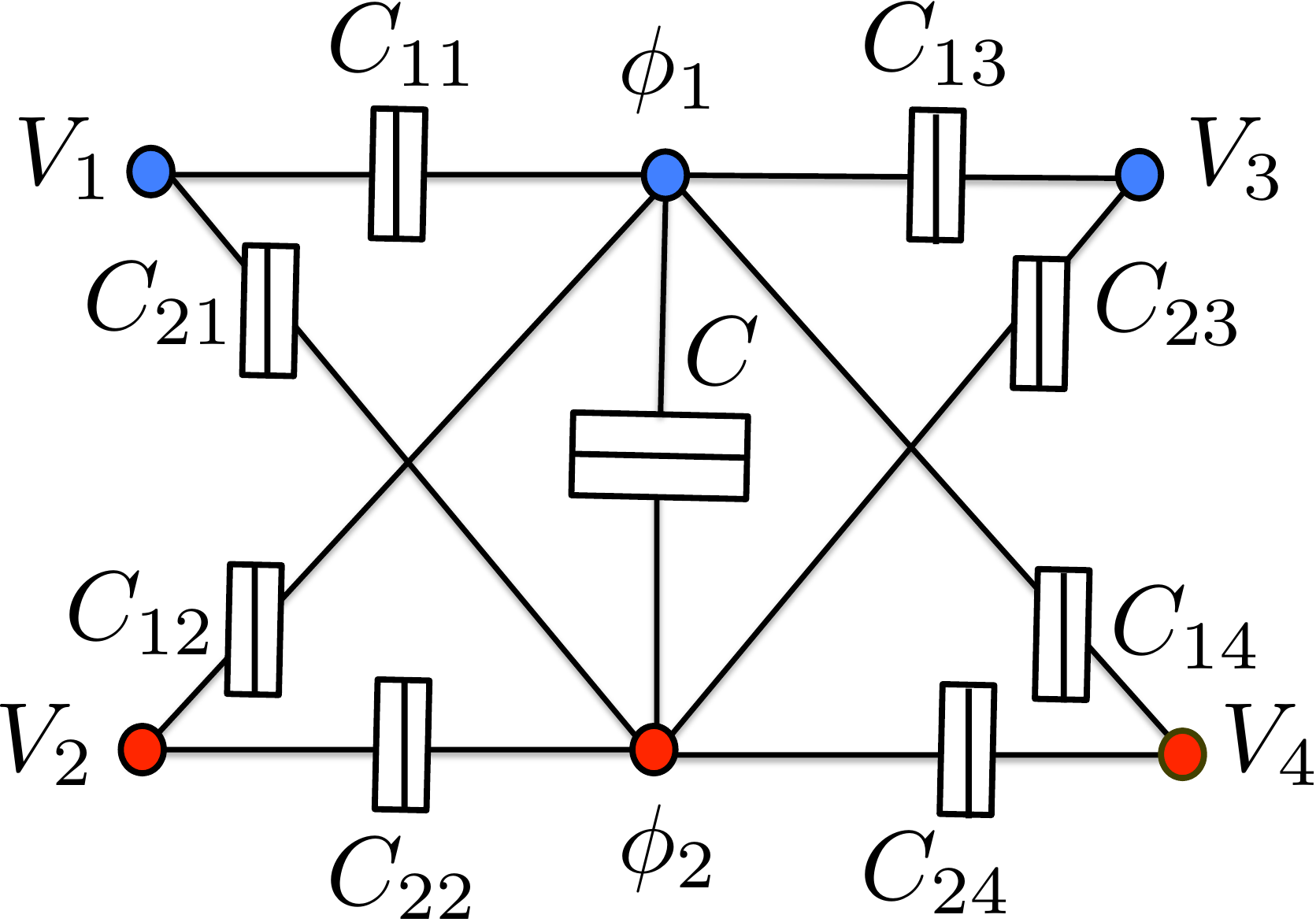}
\caption{Sketch of the electrostatic model.
\label{fig:elecmodel}}
\end{figure}

We consider two capacitively coupled quantum dots, $1$ and $2$. With the geometry shown in Supp.\ Fig.~\ref{fig:elecmodel},
the electrostatic equations for the charges $Q_1$ and $Q_2$ are given by
\begin{align}
Q_1 &= \sum_i C_{1i} (\phi_1-V_i) + C (\phi_1-\phi_2)\,,\label{eq_q1}\\
Q_2 &= \sum_i C_{2i} (\phi_2-V_i) + C (\phi_2-\phi_1)\,,\label{eq_q2}
\end{align}
with $\phi_1$ and $\phi_2$ the internal potentials and $V_i$ ($i=1,\ldots,4)$ the applied voltages.
The potential energies for both dots with
$Q_1=eN_1$ and $Q_2=eN_2$ excess electrons take the form
\begin{align}
U_1(N_1,N_2)=\int_0^{e N_1} \phi_1(Q_1,Q_2) dQ_1\,,\\
U_2(N_1,N_2)=\int_0^{e N_2} \phi_2(Q_1,Q_2) dQ_2\,,
\end{align}
where $\phi_1$ and $\phi_2$ are determined from Eqs.~\eqref{eq_q1} and~\eqref{eq_q2}.
The electrochemical potential of dots $1$ and $2$ can thus be written as
\begin{align}
\mu_1=\varepsilon_1+U_1(1,0)-U_1(0,0)\,,\\
\mu_2=\varepsilon_2+U_2(0,1)-U_d(0,0)\,.
\end{align}
We have lumped the gate dependence into the values of the dot levels $\varepsilon_1$ and $\varepsilon_2$.

Let $\mu_{li}=E_F+eV_i$ be the electrochemical potential of lead $i$ and
$K=\sum_{i} C_{1i} \sum_j C_{2j} +C \sum_{i,j} C_{ij}$.
Since the Fermi functions are evaluated at $\mu_{S1}-\mu_1$, etc.\ (see main text),
the energies of interest become
\begin{subequations}
\label{eq_eoi}
\begin{align}
\mu_1-\mu_{l1} &= \varepsilon_1-E_F+\frac{e}{K}\left[
\frac{e(C+\sum_i C_{2i})}{2}+\sum_i C_{2i} \sum_j C_{1j}V_{j1}+C\sum_{i,j} C_{ij}V_{j1}\right]\,, \\
\mu_1-\mu_{l3} &= \varepsilon_1-E_F+\frac{e}{K}\left[
\frac{e(C+\sum_i C_{2i})}{2}+\sum_i C_{2i} \sum_j C_{1j}V_{j3}+C\sum_{i,j} C_{ij}V_{j3}\right]\,, \\
\mu_2-\mu_{l2} &= \varepsilon_2-E_F+\frac{e}{K}\left[
\frac{e(C+\sum_i C_{1i})}{2}+\sum_i C_{1i} \sum_j C_{2j}V_{j2}+C\sum_{i,j} C_{ij}V_{j2}\right]\,, \\
\mu_2-\mu_{l4} &= \varepsilon_2-E_F+\frac{e}{K}\left[
\frac{e(C+\sum_i C_{1i})}{2}+\sum_i C_{1i} \sum_j C_{2j}V_{j4}+C\sum_{i,j} C_{ij}V_{j4}\right]\,, 
\end{align}
\end{subequations}
with $V_{ij}=V_i-V_j$. It is worth noting that these expressions depend on voltage differences only.
Thus, our current expression will be gauge invariant.

When one of the dots is occupied, the electrochemical potentials can be evaluated as follows,
\begin{align}
\mu_1=\varepsilon_1+\varepsilon_2+U_1(1,1)+U_2(1,1)-[\varepsilon_2+U_1(0,1)+U_2(0,1)]\,,\\
\mu_2=\varepsilon_1+\varepsilon_2+U_1(1,1)+U_2(1,1)-[\varepsilon_1+U_1(1,0)+U_2(1,0)]\,.
\end{align}
Hence, Eq.~\eqref{eq_eoi} becomes
\begin{subequations}
\label{eq_eoi2}
\begin{align}
\mu_1-\mu_{l1}&=\varepsilon_1-E_F+\frac{e}{K}\left[
\frac{5e(C+\sum_i C_{2i})}{2}+\sum_i C_{2i} \sum_j C_{1j}V_{j1}+C\sum_{i,j} C_{ij}V_{j1}\right]\,, \\
\mu_1-\mu_{l3}&=\varepsilon_1-E_F+\frac{e}{K}\left[
\frac{5e(C+\sum_i C_{2i})}{2}+\sum_i C_{2i} \sum_j C_{1j}V_{j3}+C\sum_{i,j} C_{ij}V_{j3}\right]\,, \\
\mu_2-\mu_{l2}&=\varepsilon_2-E_F+\frac{e}{K}\left[
\frac{5e(C+\sum_i C_{1i})}{2}+\sum_i C_{1i} \sum_j C_{2j}V_{j2}+C\sum_{i,j} C_{ij}V_{j2}\right]\,, \\
\mu_2-\mu_{l4}&=\varepsilon_2-E_F+\frac{e}{K}\left[
\frac{5e(C+\sum_i C_{1i})}{2}+\sum_i C_{1i} \sum_j C_{2j}V_{j4}+C\sum_{i,j} C_{ij}V_{j4}\right]\,.
\end{align}
\end{subequations}
Substracting Eq.~\eqref{eq_eoi} from Eq.~\eqref{eq_eoi2}, we find that the interdot Coulomb interaction
corresponds to
\begin{equation}\label{eq_u}
U=\frac{2e^2C}{K}\,.
\end{equation}
In our numerical calculations, we use the experimental parameters reported above:
$C_{11}=C_{S1}$, $C_{12}=C_{S2,1}$, $C_{21}=C_{S1,2}$,  $C_{22}=C_{S2}$, $C=C_m$ and
$C_{13}=C_{23}=C_{14}=C_{24}=0$ together with $V_1=V_{S1}$, $V_2=V_{S2}$, $V_3=V_{D1}$
and $V_4=V_{D2}$.

\subsection{Hamiltonian and tunnel rates}

Our model Hamiltonian is
\beq
\calh = \calh_0 + \calh_T\,,
\edq
where
\beq
\calh_0 = \calh_D + \calh_C\,.
\edq

$\calh_D$ is the Hamiltonian for the dot region,
\beq
\calh_D = \sum_{i=1,2} \veps_{i} d_i^{\dag}d_i + Un_{1}n_{2} \,,
\edq
$\calh_C$ corresponds to the reservoir Hamiltonian,
\beq
\calh_C = \sum_{\alpha=S/D,i,k} \veps_{\alpha k} c_{\alpha ik}^{\dag}c_{\alpha ik}\,,
\edq
and finally $\calh_T$ describes tunneling processes between the dot region and the reservoirs,
\beq
\calh_T = \sum_{\alpha,i,k} \left(t_{\alpha ik} c_{\alpha ik}^{\dag} d_i + h.c.\right)\,.
\edq

We regard $\calh_T$ as a perturbation and calculate the probabilities for transitions
between initial $|\psi_i\rangle$ and final $|\psi_f\rangle$ states (energies $E_i$ and $E_f$, respectively)
from an expansion of the $T$-matrix,
\beq
P_{if} = \frac{2\pi}{\hbar}\left|\nbraket{\psi_f|\calh_T + \calh_T\mathcal{G}_0\calh_T + \cdots|\psi_i}\right|^2 \delta(E_i-E_f)\,,
\edq
where the resolvent operator is ${\cal G}_0 = (E_i - \calh_0)^{-1}$.
To lowest order in $\calh_T$ (Fermi's golden rule) one obtains sequential-tunneling transition rates
between the dot states $|0\rangle=|00\rangle$, $|1\rangle=|10\rangle$,
$|2\rangle=|01\rangle$ and $|d\rangle=|11\rangle$:
\begin{align}
\Gamma_{0i}^{\alpha i} &=\frac{1}{\hbar}\Gamma_{\alpha i} f_{\alpha i}(\mu_i)\,,\\
\Gamma_{0i}^{\alpha i} &=\frac{1}{\hbar}\Gamma_{\alpha i} [1-f_{\alpha i}(\mu_i)]\,,\\
\Gamma_{id}^{\alpha\bi} &= \frac{1}{\hbar}\gamma_{\alpha\bi} f_{\alpha\bi}(\mu_{\bi}+U)\,, \\
\Gamma_{di}^{\alpha\bi} &= \frac{1}{\hbar}\gamma_{\alpha\bi} \left[1 - f_{\alpha\bi}(\mu_{\bi}+U)\right]\,,
\end{align}
with $\Gamma_{\alpha i}=2\pi \rho_{\alpha i} |t_{\alpha i}^0|^2$
and $\gamma_{\alpha i}=2\pi \rho_{\alpha i} |t_{\alpha i}^1|^2$.
As discussed in the main article, the transmission probabilities depend on energy and
$t_{\alpha i}$ thus becomes a function of the charge state. In our numerical simulations of the drag current, we use the experimental values $2\Gamma_{S1}=2\Gamma_{D1}=\Gamma_1=15$~$\mu$eV and $2\Gamma_{S2}=2\Gamma_{D2}=\Gamma_2=47$~$\mu$eV and choose $\gamma_{S1}=\gamma_{D1}=\Gamma_{S1}$, $\gamma_{S2}=0.5\Gamma_{S2}$ and $\gamma_{D2}=\Gamma_{D2}$.

To second order in $\calh_T$ we obtain cotunneling transition rates involving 
many intermediate states which must be summed over:
\beq
\gamma_{i f} = \frac{2\pi}{\hbar} \tr W_{i} \left|\langle \psi_f|\calh_T\frac{1}{E_i - \calh_0}\calh_T|\psi_i\rangle\right|^2 \delta(E_i - E_f)\,,
\edq
where the trace is performed over the lead degrees of freedom and the thermal factor $W_{i}$ obeys
\beq
\sum_{i} W_{i} \left|\langle \psi_i| c_{\alpha k}^{\dag}c_{\alpha k}|\psi_i\rangle\right| = f_{\alpha}(\veps_{\alpha k}) \,.
\edq
In the following, we focus on the nondiagonal cotunneling rates since the terms
$\gamma_{00}$, $\gamma_{11}$, $\gamma_{22}$ and $\gamma_{dd}$
do not contribute to the master equations (see below) or to the drag current:
\bes
\begin{align}
\gamma_{i\bi}^{\alpha\bi\beta i} &= \frac{2\pi}{\hbar} \int d\veps~\left|\frac{t_{\alpha\bi}^0t_{\beta i}^0}{\veps-\mu_{\bi}+i\eta} 
- \frac{t_{\alpha\bi}^1 t_{\beta i}^1}{\veps-\mu_{\bi}-U+i\eta}\right|^2
\rho_{\alpha\bi}\rho_{\beta i} f_{\alpha\bi}(\veps)\left[1 - f_{\beta i}(\veps+\mu_i-\mu_{\bi})\right]\,,\\
\gamma_{0d}^{\alpha\bi\beta i} &= \frac{2\pi}{\hbar} \int d\veps~\left|\frac{t_{\alpha\bi}^0 t_{\beta i}^1}{\veps-\mu_{\bi}+i\eta} 
- \frac{t_{\alpha\bi}^1 t_{\beta i}^0}{\veps-\mu_{\bi}-U+i\eta}\right|^2
\rho_{\alpha\bi}\rho_{\beta i} f_{\alpha\bi}(\veps)f_{\beta i}(\mu_i+\mu_{\bi}+U-\veps)\,, \\
\gamma_{d0}^{\alpha\bi\beta i} &= \frac{2\pi}{\hbar} \int d\veps~\left|\frac{t_{\alpha\bi}^0t_{\beta i}^1}{\veps-\mu_{\bi}+i\eta} 
- \frac{t_{\alpha\bi}^1 t_{\beta i}^0}{\veps-\mu_{\bi}-U+i\eta}\right|^2
\rho_{\alpha\bi}\rho_{\beta i} \left[1-f_{\alpha\bi}(\veps)\right]\left[1-f_{\beta i}(\mu_i+\mu_{\bi}+U-\veps)\right]\,,
\end{align}
\eds
where a finite broadening $\eta$ (a small imaginary part) has been added to the denominators in order to avoid the divergence associated to the infinite lifetime of the intermediate states. Expansion in powers of $1/\eta$ leads to a first term which reproduces the sequential tunneling result. Therefore, to avoid double counting we subtract this term. The next order is independent of $\eta$ and corresponds to the nondivergent cotunneling expressions. 
We find
\bes
\begin{multline}
\gamma_{i\bi}^{\alpha\bi\beta i} 
= \frac{\beta}{4\pi^2\hbar}\Gamma_{\alpha\bi}\Gamma_{\beta i}
n_B(\mu_{\bi} - \mu_i + \mu_{\beta i}-\mu_{\alpha\bi})
\Im\left\{ \Psi^{(1)}\left(\frac{1}{2} + i\beta\frac{\mu_{\bi}-\mu_{\alpha\bi}}{2\pi}\right)
-\Psi^{(1)}\left(\frac{1}{2} + i\beta\frac{\mu_i-\mu_{\beta i}}{2\pi}\right) \right\} \\
+ \frac{\beta}{4\pi^2\hbar}\gamma_{\alpha\bi}\gamma_{\beta i}
n_B(\mu_{\bi} - \mu_i + \mu_{\beta i}-\mu_{\alpha\bi})
\Im\left\{ \Psi^{(1)}\left(\frac{1}{2} + i\beta\frac{\mu_{\bi}+U-\mu_{\alpha\bi}}{2\pi}\right)
-\Psi^{(1)}\left(\frac{1}{2} + i\beta\frac{\mu_i+U-\mu_{\beta i}}{2\pi}\right) \right\}\\
-\frac{\sqrt{\Gamma_{\alpha\bi}\Gamma_{\beta i}\gamma_{\alpha\bi}\gamma_{\beta i}}}{\pi\hbar} n_B(\mu_{\bi} - \mu_i + \mu_{\beta i}-\mu_{\alpha\bi}) \frac{1}{U}
\Re\left\{ \Psi\left(\frac{1}{2} + i\beta\frac{\mu_{\bi}-\mu_{\alpha\bi}}{2\pi}\right) - \Psi\left(\frac{1}{2} - i\beta\frac{\mu_{\bi}+U-\mu_{\alpha\bi}}{2\pi}\right) \right.\\
\left. -\Psi\left(\frac{1}{2} + i\beta\frac{\mu_i-\mu_{\beta i}}{2\pi}\right) + \Psi\left(\frac{1}{2} - i\beta\frac{\mu_i+U-\mu_{\beta i}}{2\pi}\right) \right\}\,,
\end{multline}
\begin{multline}
\gamma_{0d}^{\alpha\bi\beta i} 
= \frac{\beta}{4\pi^2\hbar} \Gamma_{\alpha\bi}\gamma_{\beta i} n_B(\mu_i + \mu_{\bi} + U - \mu_{\alpha\bi} - \mu_{\beta i})
\Im\left\{ \Psi^{(1)}\left(\frac{1}{2} + i\beta\frac{\mu_{\bi}-\mu_{\alpha\bi}}{2\pi}\right)
+ \Psi^{(1)}\left(\frac{1}{2} + i\beta\frac{\mu_i+U-\mu_{\beta i}}{2\pi}\right) \right\}\\
+ \frac{\beta}{4\pi^2\hbar} \gamma_{\alpha\bi}\Gamma_{\beta i} n_B(\mu_i + \mu_{\bi} + U - \mu_{\alpha\bi} - \mu_{\beta i})
\Im\left\{ \Psi^{(1)}\left(\frac{1}{2} + i\beta\frac{\mu_{\bi}+U-\mu_{\alpha\bi}}{2\pi}\right)
+ \Psi^{(1)}\left(\frac{1}{2} + i\beta\frac{\mu_i-\mu_{\beta i}}{2\pi}\right) \right\}\\
-\frac{\sqrt{\Gamma_{\alpha\bi}\Gamma_{\beta i}\gamma_{\alpha\bi}\gamma_{\beta i}}}{\pi\hbar} n_B(\mu_i + \mu_{\bi} + U - \mu_{\alpha\bi} - \mu_{\beta i}) \frac{1}{U}
\Re\left\{ \Psi\left(\frac{1}{2} + i\beta\frac{\mu_{\bi}-\mu_{\alpha\bi}}{2\pi}\right) - \Psi\left(\frac{1}{2} - i\beta\frac{\mu_{\bi}+U-\mu_{\alpha\bi}}{2\pi}\right) \right. \\
\left. + \Psi\left(\frac{1}{2} + i\beta\frac{\mu_i-\mu_{\beta i}}{2\pi}\right) - \Psi\left(\frac{1}{2} - i\beta\frac{\mu_i+U-\mu_{\beta i}}{2\pi}\right) \right\}\,,
\end{multline}
\begin{multline}
\gamma_{d0}^{\alpha\bi\beta i} 
= \frac{\beta}{4\pi^2\hbar} \Gamma_{\alpha\bi}\gamma_{\beta i} \left[1 + n_B(\mu_i + \mu_{\bi} + U - \mu_{\alpha\bi} - \mu_{\beta i})\right]
\Im\left\{ \Psi^{(1)}\left(\frac{1}{2} + i\beta\frac{\mu_{\bi}-\mu_{\alpha\bi}}{2\pi}\right)
+ \Psi^{(1)}\left(\frac{1}{2} + i\beta\frac{\mu_i+U-\mu_{\beta i}}{2\pi}\right) \right\}\\
+ \frac{\beta}{4\pi^2\hbar} \gamma_{\alpha\bi}\Gamma_{\beta i} \left[1 + n_B(\mu_i + \mu_{\bi} + U - \mu_{\alpha\bi} - \mu_{\beta i})\right] \Im\left\{ \Psi^{(1)}\left(\frac{1}{2} + i\beta\frac{\mu_{\bi}+U-\mu_{\alpha\bi}}{2\pi}\right)
+ \Psi^{(1)}\left(\frac{1}{2} + i\beta\frac{\mu_i-\mu_{\beta i}}{2\pi}\right) \right\}\\
-\frac{\sqrt{\Gamma_{\alpha\bi}\Gamma_{\beta i}\gamma_{\alpha\bi}\gamma_{\beta i}}}{\pi\hbar} \left[1+n_B(\mu_i + \mu_{\bi} + U - \mu_{\alpha\bi} - \mu_{\beta i})\right] \frac{1}{U}
\Re\left\{ \Psi\left(\frac{1}{2} + i\beta\frac{\mu_{\bi}-\mu_{\alpha\bi}}{2\pi}\right) - \Psi\left(\frac{1}{2} - i\beta\frac{\mu_{\bi}+U-\mu_{\alpha\bi}}{2\pi}\right) 
\right. \\
\left. + \Psi\left(\frac{1}{2} + i\beta\frac{\mu_i-\mu_{\beta i}}{2\pi}\right) - \Psi\left(\frac{1}{2} - i\beta\frac{\mu_i+U-\mu_{\beta i}}{2\pi}\right) \right\}\,.
\end{multline}
\eds
Here, $n_B(x)=1/(e^{x}-1)$ is the Bose distribution function, $\Psi$ ($\Psi^{(1)}$) is the digamma (trigamma)
function and $\beta=1/k_B T$ is the inverse temperature.

The stationary values of the set of probabilities $\boldsymbol{p}=(p_0,p_1,p_2,p_d)^T$
follow from the equations $0=\mathbf{\Gamma} \boldsymbol{p}$ written in matrix form with
\begin{equation}
\mathbf{\Gamma}=\begin{pmatrix}
-\left(\Gamma_{01} + \Gamma_{02} + \gamma_{0d}\right) & \Gamma_{10} & \Gamma_{20} & \gamma_{d0} \\
\Gamma_{01} & -\left(\Gamma_{10} + \Gamma_{1d} + \gamma_{12}\right) & \gamma_{21} & \Gamma_{d1} \\
\Gamma_{02} & \gamma_{12} & - \left(\Gamma_{20} + \Gamma_{2d} + \gamma_{21}\right) & \Gamma_{d2} \\
\gamma_{0d} & \Gamma_{1d} & \Gamma_{2d} & - \left(\Gamma_{d1} + \Gamma_{d2} + \gamma_{d0}\right) \,,
\end{pmatrix}
\end{equation}
where
\bes
\beq
\begin{split}
\Gamma_{01} &= \sum_{\alpha} \Gamma_{01}^{\alpha 1}, \quad
\Gamma_{10} = \sum_{\alpha} \Gamma_{10}^{\alpha 1}, \quad
\Gamma_{02} = \sum_{\alpha} \Gamma_{02}^{\alpha 2}, \quad
\Gamma_{20} = \sum_{\alpha} \Gamma_{20}^{\alpha 2},
\\
\Gamma_{1d} &= \sum_{\alpha} \Gamma_{1d}^{\alpha 2}, \quad
\Gamma_{d1} = \sum_{\alpha} \Gamma_{d1}^{\alpha 2}, \quad
\Gamma_{2d} = \sum_{\alpha} \Gamma_{2d}^{\alpha 1}, \quad
\Gamma_{d2} = \sum_{\alpha} \Gamma_{d2}^{\alpha 1},
\\
\gamma_{12} &= \sum_{\alpha,\beta} \gamma_{12}^{\alpha 2\beta 1}, \quad
\gamma_{21} = \sum_{\alpha,\beta} \gamma_{21}^{\alpha 1\beta 2}, \quad
\gamma_{0d} = \sum_{\alpha,\beta} \gamma_{0d}^{\alpha 1\beta 2}, \quad
\gamma_{d0} = \sum_{\alpha,\beta} \gamma_{d0}^{\alpha 1\beta 2},
\end{split}
\edq
\eds

\newpage
\section{Extended data and analysis from Fig. 2}

\begin{figure}
\includegraphics{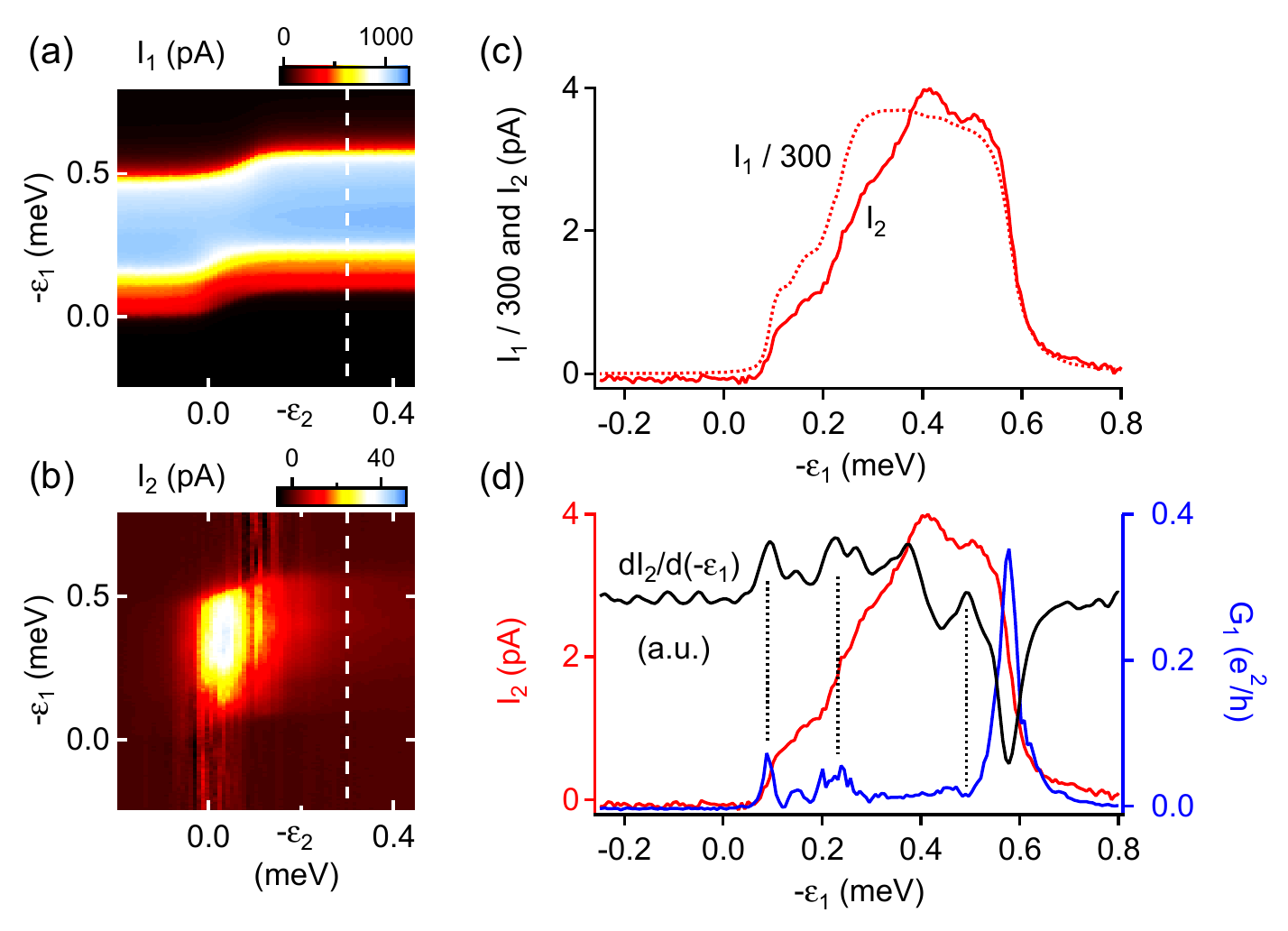}
\caption{Extended data and analysis from Fig. 2 ($V_{S1} = 0.5$~mV, $V_{S2} = 0$; $-\varepsilon_2 = 0.3$~meV where cuts are shown). (a) Drive current $I_1$. (b)~Drag current $I_2$. This is a reproduction of Fig. 2(e) to aid in comparing with \subfig{cutcomp}{a}. (c)~Cuts from \subfig{cutcomp}{a} and \subfig{cutcomp}{b}. $I_1$ has been divided by 300 to fit on the same scale. (d)~Numerically differentiated $dI_2/d(-\epsilon_1)$ (black, arbitrary units) is correlated, positively or negatively, with $G_1$ (blue, right axis). $G_1$ has been taken from the cut in Fig. 2(c) and $I_2$ is the same trace as above in \subfig{cutcomp}{c}. The correlated features appear generic in region (iii) as defined in Fig. 2(a).
\label{fig:cutcomp}}
\end{figure}

In \subfig{cutcomp}{a} we show the drive current $I_1$ as a function of $-\varepsilon_1$ and $-\varepsilon_2$ for the same parameters as used in Fig. 2. For ease of comparison, \subfig{cutcomp}{b} reproduces Fig. 2(e), the drag current $I_2$. \Startsubfig{cutcomp}{c} compares $I_1$ and $I_2$ for $-\varepsilon_2 = 0.3$~meV. The drag current is seen to be less than a percent of the drive current, which is typical for region (iii), although the ratio can reach a few percent in region (ii).

In \allfig{cutcomp}{d} we compare the vertical cuts in Fig. 2(c) and 2(e). Peaks and dips in $G_1$ are correlated (or anticorrelated) with peaks and dips in numerically calculated $dI_2/d(-\varepsilon_1)$. There appears to be correlation near $-\varepsilon_1 = U = 0.1$~meV, and anticorrelation nearer to $-\varepsilon_1 = U + |e|V_{S1} = 0.6$~meV. In the middle, $G_1$ is flat and there appears to be no correlation. As the peaks in $G_1$ correspond to excited states in dot 1, these features may not be explained satisfactorily by existing theories, and are well-resolved in the low temperature regime.

\bibliography{supp}